\documentclass[fleqn,usenatbib]{mnras}
\usepackage{txfonts}

\usepackage[T1]{fontenc}
\usepackage{ae,aecompl, longtable, lscape}

\usepackage{color, soul, ulem,graphicx, amsbsy}

\newcommand{\msano}{{\rm M}_\odot~{\rm yr}^{-1}}
\newcommand{\e}[1]{\times 10^{#1}}

\title[Can we detect aurora in exoplanets orbiting M dwarfs?]{Can we detect aurora in exoplanets orbiting M dwarfs?}
\author[Vidotto, Feeney \& Groh]{A. A. Vidotto$^{1}$, N. Feeney$^{1}$, J.~H.~Groh$^{1}$\thanks{E-mail: aline.vidotto@tcd.ie}
\\
$^{1}$ School of Physics, Trinity College Dublin, the University of Dublin, Dublin-2, Ireland
}

\date{Accepted XXX. Received YYY; in original form ZZZ}

\pubyear{}

\begin{document}
\label{firstpage}
\pagerange{\pageref{firstpage}--\pageref{lastpage}}
\maketitle

\begin{abstract}
New instruments and telescopes, such as SPIRou, CARMENES and TESS, will increase manyfold the number of known planets orbiting M dwarfs. To guide future radio observations, we estimate radio emission from known M-dwarf planets using the empirical radiometric prescription derived in the solar system, in which radio emission is powered by the wind of the host star.  Using solar-like  wind models, we find that the most promising exoplanets for radio detections are GJ 674 b and Proxima b, followed by YZ Cet b, GJ 1214 b, GJ 436 b. These are the systems that are the closest to us ($<10$ pc). However, we also show that our radio fluxes are very sensitive to the unknown properties of  winds  of M dwarfs. So, which types of winds would generate detectable radio emission? In a `reverse engineering' calculation, we show that winds with mass-loss rates $\dot{M} \gtrsim \kappa_{\rm sw} /u_{\rm sw}^3$ would drive  planetary radio emission detectable with present-day instruments, where $u_{\rm sw}$ is the local stellar wind velocity and $\kappa_{\rm sw}$ is a constant that depends on the size of the planet, distance and orbital radius. Using observationally-constrained properties of the quiescent winds of GJ~436 and Proxima Cen, we conclude that it is unlikely that GJ 436 b and Proxima b would be detectable with present-day radio instruments, unless the host stars generate episodic coronal mass ejections. GJ 674 b, GJ 876 b and YZ Cet b could present good prospects for radio detection, provided that their host-stars' winds have $\dot{M} u_{\rm sw}^{3} \gtrsim 1.8\e{-4}  \msano ({\rm km/s})^{3} $. 
\end{abstract}
\begin{keywords}
stars: planetary systems -- stars: low-mass -- stars: winds, outflows -- planet-star interactions 
\end{keywords}

%
\section{Introduction}\label{sec.intro}
The search for and characterisation of exoplanets is central to a number of current astrophysical research questions and space exploration. There has been rapid progress in exoplanet detection in recent years, with now over 3,800 confirmed exoplanets on record. The most successful detection methods so far have been the transit and the radial velocity methods. Both of these methods are indirect methods of exoplanet detection, which infer the existence of an exoplanet from the effect it has on the host star. 

Exoplanets can  be directly detected through imaging. A limitation of such method is the high contrast ratio between the intensity of electromagnetic radiation of the planet and its host star in the visible and infrared ranges; approximately 10\textsuperscript{9} in the visible and 10\textsuperscript{6} in the infrared \citep{2007P&SS...55..598Z}. However, there may exist a direct detection method in the low frequency radio range, between ten and a few hundred MHz \citep{2011RaSc...46.0F09G}, given that some planets in the solar system are known to emit radiation within this range. This radiation arises via the interaction between planetary magnetic fields and the solar wind \citep{DeschKaiser84} and is known as auroral radio emission. For all magnetised solar system planets, this radiation is only 1 to 2 orders of magnitude less intense than the radiation produced by the Sun in the low frequency radio range \citep{2007P&SS...55..598Z}, thus making it favourable for direct detection. A nearly linear relation between the emitted radio power of these planets and the dissipated kinetic power of the incident solar wind has been observed \citep{DeschKaiser84} and is known as the `radiometric Bode's law' . 

It has been theorised that magnetised exoplanets may emit at radio frequencies, analogously to the solar system planets \citep[e.g.,][]{1999JGR...10414025F, 2004ApJ...612..511L}. If the radiometric Bode's law holds true for exoplanetary systems in a similar way as for the solar system planets, one can use stellar wind models to estimate radio emission from exoplanets. It is expected that hot Jupiters produce radio emission that are many orders of magnitude larger than Jupiter \citep[][but see also \citealt{2017MNRAS.469.3505W, 2019MNRAS.485.4529K}]{2005A&A...437..717G, 2007A&A...475..359G, 2010ApJ...720.1262V, 2012MNRAS.423.3285V}, the strongest emitter in the solar system at radio wavelength. This is because hot Jupiters orbit at regions where the stellar wind has a large ram pressure, which can power stronger planetary radio emissions. 

The detection of exoplanetary radio emissions would not only be a revolutionary method of exoplanet detection, but also an indicative of the presence of an intrinsic planetary magnetic field. Although some studies have proposed the presence of exoplanetary magnetism to interpret spectroscopic transit observations \citep[e.g.,][]{2010ApJ...722L.168V, 2011MNRAS.416L..41L, 2014Sci...346..981K}, the interpretations are not unique \citep{2015ASSL..411..153V} and there is still no conclusive detection of an exoplanet magnetic field. Planetary magnetic fields are believed to be one of the key ingredients  for determining planetary habitability \citep[e.g.][]{2007AsBio...7..185L, 2019MNRAS.485.3999M}. 

In that regard, M dwarf stars have been the prime targets for detecting terrestrial planets in the habitable zone, which is the region around a star in which a planet could host liquid water on its surface \citep{1993Icar..101..108K}. An important issue affecting habitability of M dwarf planets is that their host stars remain active for a long part of their lives \citep{2008AJ....135..785W,2011ApJ...727...56I} and have high flare and coronal mass ejections (CME) rates \citep{2014ApJ...797..122D, 2016A&A...590A..11V,2017ApJ...841..124V, 2016ApJ...826..195K}. This intense activity could be dangerous for a habitable zone planet, as these planets would receive high dosages of high-energy radiation and intense stellar wind/CME, which could strip away their atmospheres \citep{2007AsBio...7..185L, 2007AsBio...7..167K, 2007AsBio...7...85S, 2013A&A...557A..67V}. Although intense winds could be dangerous for the survival of planetary atmospheres, intense winds and CMEs would  power stronger exoplanetary radio emission, thus allowing one to probe magnetism in exoplanets.

Recently, \citet{2017ApJ...849L..10B} predicted the radio emission of the closest exoplanet to us -- Proxima b, which orbits an M dwarf star. They estimate Proxima b has radio emission as high as 1~Jy, in a frequency of 0.02 MHz. Due to its close distance to us, Proxima b is one of the most likely candidates for producing radio emissions that could be detectable from Earth (although, with such predicted frequencies, this would not be observed with ground-based instrumentation). In this work, we estimate the radio emission of exoplanets orbiting M-dwarf stars (for predictions relating to hot Jupiters around solar-type stars see, for example, \citealt{2007A&A...475..359G, 2011RaSc...46.0F09G}). At the time of writing, there are nearly 200 planets orbiting stars in the mass range between 0.1 and 0.5~$M_\odot$. The number of known exoplanets orbiting M dwarfs, including exoplanets in their habitable zones, will increase manyfold, with surveys conducted by, e.g., SPIRou \citep{2018AJ....155...93C}, CARMENES \citep{2014SPIE.9147E..1FQ} and TESS  \citep{2015ApJ...809...77S}.  Due to detection bias, most of the known M-dwarf planets have small semi-major axis, making them favourable candidates for detecting exoplanetary radio emissions. 

This paper is divided as follows. Section \ref{sec.candidate} presents our candidate selection and Section \ref{sec.model} shows the radio emission model, which is based in the work of \citet{2017A&A...602A..39V}. Section \ref{sec.results} shows our results, where we estimate the radio emission of exoplanets using two different approaches. In the first approach (Section \ref{sec.app1}), we use simple stellar wind models to derive radio fluxes, and we demonstrate the large influence that stellar wind models have on the predicted radio emission from exoplanets. Since stellar wind properties of M dwarfs are uncertain, in Section \ref{sec.app2} we present a second approach based on a `reverse engineering' investigation,  in which we derive what are the stellar wind properties that would power {\it detectable} radio emission with present-day instrumentation. This is then followed by our discussions (Section \ref{sec.discussion}) and conclusions (Section \ref{sec.conclusions}).

\section{Candidate selection}\label{sec.candidate}
We make use of the NASA Exoplanet Archive\footnote{\url{https://exoplanetarchive.ipac.caltech.edu/}} to select 120 exoplanets  (Table \ref{table_physics}) that fulfil the following selection criteria.
\begin{enumerate}
\item We select stars in the mass range between 0.1 and 0.5$M_\odot$. Some of the selected stars do not have quoted errors in their masses (namely, KOI-55, Wolf 1061, GJ 3323, GJ 3341, KOI-55, HD 285968, GJ 674, Ross 458, GJ 273, GJ 3293, GJ 3323). For all the others, the errors quoted in stellar masses are within $25\%$, with the exception of the following stars:  K2-72, Kepler-42, K2-9, BD-08 2823, HD 125595, HIP 57274, HD 99492. For the last three stars, masses are quite uncertain ($\sim$ a factor of 2). 
\item Kepler's third law was used to convert from orbital period to semi-major axis in the case of planets without defined semi-major axis values on the catalogue. 
\item Given that the planetary radius is necessary for the computation of planetary radio power and flux, we use the analytical expression from \citet[][see their equation (23)]{2007ApJ...669.1279S} to derive the radii of planets without quoted radii (mostly  planets that are not transiting).  These planets are highlighted in Table \ref{table_physics} with an asterisk. This analytical expression offers an estimate of planetary radii and is valid for planets with masses below $\sim 23~M_\oplus$. We use the formalism for planets with Fe composition, which gives a {\it lower limit} on the size of rocky planets \citep{2007ApJ...669.1279S}. Since the square of the planet radius  $R_p$ enters in the radio power computation, a {\it lower limit} on $R_p$ produces a conservative estimate for planetary radio power and flux. Assuming a silicate composition (MgSiO$_3$), for example, would increase the quoted radii of these planets by a factor of $1.29$ to $1.35$, which would increase our estimated radio power by a factor of up to $1.8$. Planets with masses above the $\sim 23~M_\oplus$ limit or without quoted masses and radii were not included in our sample. 
\item Finally, some stars did not have quoted radii in the catalogue. For the following stars, we used the radii quoted in \citet{2001A&A...367..521P}: GJ 667 C, GJ 163, GJ 674 and GJ 832. 
\end{enumerate}

\begin{table*}
\caption{Sample of exoplanets studied in this work. Data are from the NASA Exoplanet Archive for most of the properties. Planets  with estimated radii assuming Fe composition are highlighted with an asterisk. See Section \ref{sec.candidate} for a detailed explanation. The last column is the wind constant $\kappa_{\rm sw}$ that obeys the relation $\dot{M} \gtrsim \kappa_{\rm sw} /u_{\rm sw}^3$, with mass-loss rate $\dot{M}$ given in $\msano$ and local stellar wind velocity $u_{\rm sw}$ in km/s. This equation provides the minimum mass-loss rates required for a stellar wind to produce detectable radio emission with present-day instruments ($> 1$~mJy, cf.~Section \ref{sec.app2}).} 
\label{table_physics}
\begin{center}
\begin{tabular}{lccccccccccccc}
\hline
Planet &$M_\star$&$R_\star$  & $d$ & $M_p$ & $R_p$ & $\langle \rho_p \rangle$&$P_{\rm orb}$ &sma & sma  & $\kappa_{\rm sw}$  \\
name &  ($M_\odot$)&  ($R_\odot$)& (pc)&($M_\oplus$) &($R_\oplus$) & (g cm$^{-3}$) & (d) & (au)& ($R_\star$)  & ($M_\odot/{\rm yr \, (km/s)}^{3}$)\\
\hline \hline
Proxima b$^*$ &$  0.12 $&$  0.14 $&$ 1.29 $&$ 1.3 $&$ 1.3 $&$ 3.1 $&$ 11 $&$ 
     0.049 $&$ 74.5 $&$ 1.56\e{-5} $ \\ 
GJ 876 d$^*$ &$  0.32 $&$  0.30 $&$ 4.68 $&$ 6.7 $&$ 2.1 $&$ 4.1 $&$ 1.9 $&$ 
     0.021 $&$ 14.9 $&$ 7.92\e{-5} $ \\ 
GJ 674 b$^*$ &$  0.35 $&$  0.43 $&$ 4.55 $&$ 11 $&$ 2.3 $&$ 4.7 $&$ 4.7 $&$ 
     0.039 $&$ 19.5 $&$ 1.50\e{-4} $ \\ 
GJ 436 b &$  0.47 $&$  0.46 $&$ 9.76 $&$ 22 $&$ 4.2 $&$ 1.7 $&$ 2.6 $&$ 
     0.029 $&$ 13.6 $&$ 1.78\e{-4} $ \\ 
YZ Cet b$^*$ &$  0.13 $&$  0.17 $&$ 3.60 $&$ 0.75 $&$ 1.1 $&$ 3.0 $&$ 2.0 $&$ 
     0.016 $&$ 19.7 $&$ 1.82\e{-4} $ \\ 
GJ 411 b$^*$ &$  0.39 $&$  0.39 $&$ 2.55 $&$ 3.0 $&$ 1.7 $&$ 3.5 $&$ 13 $&$ 
     0.079 $&$ 43.3 $&$ 2.34\e{-4} $ \\ 
YZ Cet c$^*$ &$  0.13 $&$  0.17 $&$ 3.60 $&$ 0.98 $&$ 1.2 $&$ 3.0 $&$ 3.1 $&$ 
     0.021 $&$ 26.4 $&$ 2.39\e{-4} $ \\ 
YZ Cet d$^*$ &$  0.13 $&$  0.17 $&$ 3.60 $&$ 1.1 $&$ 1.3 $&$ 3.1 $&$ 4.7 $&$ 
     0.028 $&$ 34.9 $&$ 3.48\e{-4} $ \\ 
GJ 581 b$^*$ &$  0.31 $&$  0.29 $&$ 6.30 $&$ 16 $&$ 2.5 $&$ 5.2 $&$ 5.4 $&$ 
     0.041 $&$ 30.1 $&$ 4.31\e{-4} $ \\ 
GJ 1214 b &$  0.15 $&$  0.22 $&$ 12.9 $&$ 6.3 $&$ 2.8 $&$ 1.5 $&$ 1.6 $&$ 
     0.014 $&$ 13.8 $&$ 5.96\e{-4} $ \\ 
Ross 128 b$^*$ &$  0.17 $&$  0.20 $&$ 3.38 $&$ 1.4 $&$ 1.3 $&$ 3.2 $&$ 9.9 $&$ 
     0.050 $&$ 53.3 $&$ 6.86\e{-4} $ \\ 
Wolf 1061 b$^*$ &$  0.29 $&$  0.31 $&$ 4.31 $&$ 1.9 $&$ 1.5 $&$ 3.3 $&$ 4.9 $&$ 
     0.037 $&$ 26.0 $&$ 7.23\e{-4} $ \\ 
GJ 273 c$^*$ &$  0.29 $&$  0.29 $&$ 3.80 $&$ 1.2 $&$ 1.3 $&$ 3.1 $&$ 4.7 $&$ 
     0.036 $&$ 27.0 $&$ 7.24\e{-4} $ \\ 
GJ 3323 b$^*$ &$  0.16 $&$  0.12 $&$ 5.32 $&$ 2.0 $&$ 1.5 $&$ 3.3 $&$ 5.4 $&$ 
     0.033 $&$ 58.8 $&$ 1.21\e{-3} $ \\ 
GJ 273 b$^*$ &$  0.29 $&$  0.29 $&$ 3.80 $&$ 2.9 $&$ 1.7 $&$ 3.5 $&$ 19 $&$ 
     0.091 $&$ 67.5 $&$ 1.62\e{-3} $ \\ 
GJ 687 b$^*$ &$  0.45 $&$  0.43 $&$ 4.55 $&$ 19 $&$ 2.7 $&$ 5.6 $&$ 38 $&$ 
     0.170 $&$ 85.0 $&$ 1.74\e{-3} $ \\ 
Wolf 1061 c$^*$ &$  0.29 $&$  0.31 $&$ 4.31 $&$ 3.4 $&$ 1.7 $&$ 3.6 $&$ 18 $&$ 
     0.089 $&$ 61.7 $&$ 2.13\e{-3} $ \\ 
GJ 581 e$^*$ &$  0.31 $&$  0.29 $&$ 6.30 $&$ 1.7 $&$ 1.4 $&$ 3.2 $&$ 3.1 $&$ 
     0.028 $&$ 20.9 $&$ 2.15\e{-3} $ \\ 
GJ 667 C b$^*$ &$  0.33 $&$  0.42 $&$ 6.80 $&$ 5.7 $&$ 2.0 $&$ 4.0 $&$ 7.2 $&$ 
     0.051 $&$ 25.8 $&$ 2.44\e{-3} $ \\ 
GJ 1265 b$^*$ &$  0.18 $&$  0.19 $&$ 10.3 $&$ 7.4 $&$ 2.1 $&$ 4.2 $&$ 3.7 $&$ 
     0.026 $&$ 29.4 $&$ 2.57\e{-3} $ \\ 
GJ 581 c$^*$ &$  0.31 $&$  0.29 $&$ 6.30 $&$ 5.5 $&$ 2.0 $&$ 4.0 $&$ 13 $&$ 
     0.072 $&$ 53.4 $&$ 3.82\e{-3} $ \\ 
LHS 3844 b &$  0.15 $&$  0.19 $&$ 14.9 $&$  -  $&$ 1.3 $&$  -  $&$ 0.46 $&$ 
     0.006 $&$ 7.04 $&$ 4.67\e{-3} $ \\ 
GJ 3779 b$^*$ &$  0.27 $&$  0.28 $&$ 13.7 $&$ 8.0 $&$ 2.2 $&$ 4.3 $&$ 3.0 $&$ 
     0.026 $&$ 20.0 $&$ 7.67\e{-3} $ \\ 
GJ 832 c$^*$ &$  0.45 $&$  0.48 $&$ 4.97 $&$ 5.4 $&$ 2.0 $&$ 4.0 $&$ 36 $&$ 
     0.163 $&$ 73.0 $&$ 7.69\e{-3} $ \\ 
Kapteyn c$^*$ &$  0.28 $&$  0.29 $&$ 3.93 $&$ 7.0 $&$ 2.1 $&$ 4.2 $&$ 120 $&$ 
     0.311 $&$ 231 $&$ 8.39\e{-3} $ \\ 
GJ 1132 b$^*$ &$  0.18 $&$  0.21 $&$ 12.0 $&$ 1.7 $&$ 1.4 $&$ 3.2 $&$ 1.6 $&$ 
     0.015 $&$ 15.7 $&$ 8.62\e{-3} $ \\ 
GJ 876 e$^*$ &$  0.32 $&$  0.30 $&$ 4.68 $&$ 15 $&$ 2.5 $&$ 5.1 $&$ 120 $&$ 
     0.334 $&$ 240 $&$ 9.54\e{-3} $ \\ 
GJ 625 b$^*$ &$  0.30 $&$  0.31 $&$ 6.49 $&$ 2.8 $&$ 1.6 $&$ 3.5 $&$ 15 $&$ 
     0.078 $&$ 54.3 $&$ 1.05\e{-2} $ \\ 
HD 285968 b$^*$ &$  0.45 $&$  0.45 $&$ 9.47 $&$ 8.3 $&$ 2.2 $&$ 4.4 $&$ 8.8 $&$ 
     0.066 $&$ 31.5 $&$ 1.08\e{-2} $ \\ 
Gl 686 b$^*$ &$  0.42 $&$  0.42 $&$ 8.16 $&$ 7.1 $&$ 2.1 $&$ 4.2 $&$ 16 $&$ 
     0.091 $&$ 46.6 $&$ 1.31\e{-2} $ \\ 
GJ 3323 c$^*$ &$  0.16 $&$  0.12 $&$ 5.32 $&$ 2.3 $&$ 1.6 $&$ 3.4 $&$ 41 $&$ 
     0.126 $&$ 226 $&$ 1.54\e{-2} $ \\ 
GJ 667 C c$^*$ &$  0.33 $&$  0.42 $&$ 6.80 $&$ 3.8 $&$ 1.8 $&$ 3.7 $&$ 28 $&$ 
     0.125 $&$ 64.0 $&$ 2.30\e{-2} $ \\ 
Wolf 1061 d$^*$ &$  0.29 $&$  0.31 $&$ 4.31 $&$ 7.7 $&$ 2.1 $&$ 4.3 $&$ 220 $&$ 
     0.470 $&$ 326 $&$ 2.51\e{-2} $ \\ 
GJ 3634 b$^*$ &$  0.45 $&$  0.43 $&$ 19.8 $&$ 8.3 $&$ 2.2 $&$ 4.4 $&$ 2.6 $&$ 
     0.029 $&$ 14.3 $&$ 3.89\e{-2} $ \\ 
LHS 1140 c &$  0.18 $&$  0.21 $&$ 12.5 $&$ 1.8 $&$ 1.3 $&$ 4.8 $&$ 3.8 $&$ 
     0.027 $&$ 27.4 $&$ 4.54\e{-2} $ \\ 
GJ 163 b$^*$ &$  0.40 $&$  0.44 $&$ 15.1 $&$ 11 $&$ 2.3 $&$ 4.7 $&$ 8.6 $&$ 
     0.061 $&$ 29.7 $&$ 4.66\e{-2} $ \\ 
GJ 1132 c$^*$ &$  0.18 $&$  0.21 $&$ 12.0 $&$ 2.6 $&$ 1.6 $&$ 3.5 $&$ 8.9 $&$ 
     0.048 $&$ 48.7 $&$ 4.92\e{-2} $ \\ 
GJ 667 C f$^*$ &$  0.33 $&$  0.42 $&$ 6.80 $&$ 2.5 $&$ 1.6 $&$ 3.4 $&$ 39 $&$ 
     0.156 $&$ 79.8 $&$ 5.61\e{-2} $ \\ 
GJ 3998 b$^*$ &$  0.50 $&$  0.49 $&$ 17.8 $&$ 2.5 $&$ 1.6 $&$ 3.4 $&$ 2.6 $&$ 
     0.029 $&$ 12.7 $&$ 9.38\e{-2} $ \\ 
GJ 667 C e$^*$ &$  0.33 $&$  0.42 $&$ 6.80 $&$ 2.5 $&$ 1.6 $&$ 3.4 $&$ 62 $&$ 
     0.213 $&$ 109 $&$ 1.05\e{-1} $ \\ 
LHS 1140 b &$  0.18 $&$  0.21 $&$ 12.5 $&$ 7.0 $&$ 1.7 $&$ 7.5 $&$ 25 $&$ 
     0.094 $&$ 95.8 $&$ 1.67\e{-1} $ \\ 
K2-25 b &$  0.29 $&$  0.29 $&$ 45.0 $&$  -  $&$ 3.4 $&$  -  $&$ 3.5 $&$ 
     0.030 $&$ 22.1 $&$ 1.84\e{-1} $ \\ 
GJ 4276 b$^*$ &$  0.41 $&$  0.41 $&$ 21.3 $&$ 17 $&$ 2.6 $&$ 5.3 $&$ 13 $&$ 
     0.082 $&$ 43.0 $&$ 2.22\e{-1} $ \\ 
GJ 163 c$^*$ &$  0.40 $&$  0.44 $&$ 15.1 $&$ 6.8 $&$ 2.1 $&$ 4.2 $&$ 26 $&$ 
     0.125 $&$ 61.3 $&$ 3.09\e{-1} $ \\ 
GJ 3998 c$^*$ &$  0.50 $&$  0.49 $&$ 17.8 $&$ 6.3 $&$ 2.0 $&$ 4.1 $&$ 14 $&$ 
     0.089 $&$ 39.0 $&$ 3.23\e{-1} $ \\ 
GJ 667 C g$^*$ &$  0.33 $&$  0.42 $&$ 6.80 $&$ 4.5 $&$ 1.9 $&$ 3.8 $&$ 260 $&$ 
     0.549 $&$ 281 $&$ 3.76\e{-1} $ \\ 
GJ 3341 b$^*$ &$  0.47 $&$  0.44 $&$ 23.2 $&$ 6.6 $&$ 2.1 $&$ 4.1 $&$ 14 $&$ 
     0.089 $&$ 43.5 $&$ 8.77\e{-1} $ \\ 
HIP 57274 b$^*$ &$  0.29 $&$  0.78 $&$ 25.9 $&$ 6.4 $&$ 2.0 $&$ 4.1 $&$ 8.1 $&$ 
     0.070 $&$ 19.3 $&$ 8.81\e{-1} $ \\ 
GJ 3293 e$^*$ &$  0.42 $&$  0.40 $&$ 20.2 $&$ 3.3 $&$ 1.7 $&$ 3.6 $&$ 13 $&$ 
     0.082 $&$ 44.1 $&$ 9.11\e{-1} $ \\ 
HD 85512 b$^*$ &$  0.43 $&$  0.71 $&$ 11.3 $&$ 3.2 $&$ 1.7 $&$ 3.6 $&$ 58 $&$ 
     0.260 $&$ 78.7 $&$ 9.20\e{-1} $ \\ 
K2-18 c$^*$ &$  0.36 $&$  0.41 $&$ 34.0 $&$ 7.5 $&$ 2.1 $&$ 4.3 $&$ 9.0 $&$ 
     0.060 $&$ 31.5 $&$ 1.63\e{0} $ \\ 
HD 125595 b$^*$ &$  0.29 $&$  0.74 $&$ 28.2 $&$ 6.4 $&$ 2.0 $&$ 4.1 $&$ 9.7 $&$ 
     0.080 $&$ 23.2 $&$ 1.63\e{0} $ \\ 
K2-28 b &$  0.26 $&$  0.29 $&$ 63.1 $&$  -  $&$ 2.3 $&$  -  $&$ 2.3 $&$ 
     0.021 $&$ 15.9 $&$ 1.75\e{0} $ \\ 
Kepler-42 c &$  0.13 $&$  0.17 $&$ 38.7 $&$  -  $&$ 0.73 $&$  -  $&$ 0.45 $&$ 
     0.006 $&$ 7.59 $&$ 2.00\e{0} $ \\ 
GJ 3293 d$^*$ &$  0.42 $&$  0.40 $&$ 20.2 $&$ 7.6 $&$ 2.1 $&$ 4.3 $&$ 48 $&$ 
     0.194 $&$ 104 $&$ 2.09\e{0} $ \\ 
BD-08 2823 b$^*$ &$  0.50 $&$  0.71 $&$ 41.4 $&$ 13 $&$ 2.4 $&$ 4.9 $&$ 5.6 $&$ 
     0.060 $&$ 18.2 $&$ 2.14\e{0} $ \\ 
GJ 3293 c$^*$ &$  0.42 $&$  0.40 $&$ 20.2 $&$ 21 $&$ 2.7 $&$ 5.8 $&$ 120 $&$ 
     0.362 $&$ 194 $&$ 2.81\e{0} $ \\ 
Kepler-42 b &$  0.13 $&$  0.17 $&$ 38.7 $&$  -  $&$ 0.78 $&$  -  $&$ 1.2 $&$ 
     0.012 $&$ 14.7 $&$ 5.57\e{0} $ \\ 

 \hline
\end{tabular}
\end{center}
\end{table*}
\begin{table*}
\contcaption{} 
\begin{center}
\begin{tabular}{lccccccccccccc}
\hline
Planet &$M_\star$&$R_\star$  & $d$ & $M_p$ & $R_p$ & $\langle \rho_p \rangle$&$P_{\rm orb}$ &sma & sma  & $\kappa_{\rm sw}$  \\
name &  ($M_\odot$)&  ($R_\odot$)& (pc)&($M_\oplus$) &($R_\oplus$) & (g cm$^{-3}$) & (d) & (au)& ($R_\star$)  & ($M_\odot/{\rm yr \, (km/s)}^{3}$)\\
\hline \hline
K2-18 b &$  0.36 $&$  0.41 $&$ 34.0 $&$ 8.9 $&$ 2.4 $&$ 3.7 $&$ 33 $&$ 
     0.143 $&$ 74.9 $&$ 6.10\e{0} $ \\ 
K2-146 b &$  0.33 $&$  0.33 $&$ 79.5 $&$  -  $&$ 2.2 $&$  -  $&$ 2.6 $&$ 
     0.026 $&$ 16.9 $&$ 8.74\e{0} $ \\ 
K2-129 b &$  0.36 $&$  0.36 $&$ 27.8 $&$  -  $&$ 1.0 $&$  -  $&$ 8.2 $&$ 
     0.057 $&$ 34.0 $&$ 1.15\e{1} $ \\ 
Kepler-1624 b &$  0.50 $&$  0.47 $&$ 199 $&$  -  $&$ 5.7 $&$  -  $&$ 3.3 $&$ 
     0.034 $&$ 15.7 $&$ 1.22\e{1} $ \\ 
Kepler-445 c &$  0.18 $&$  0.21 $&$ 90.0 $&$  -  $&$ 2.5 $&$  -  $&$ 4.9 $&$ 
     0.032 $&$ 32.5 $&$ 1.25\e{1} $ \\ 
HD 99492 b$^*$ &$  0.48 $&$  0.83 $&$ 55.7 $&$ 22 $&$ 2.7 $&$ 5.9 $&$ 17 $&$ 
     0.120 $&$ 31.1 $&$ 1.71\e{1} $ \\ 
K2-91 b &$  0.29 $&$  0.28 $&$ 62.7 $&$  -  $&$ 1.1 $&$  -  $&$ 1.4 $&$ 
     0.016 $&$ 12.6 $&$ 2.00\e{1} $ \\ 
Kepler-42 d &$  0.13 $&$  0.17 $&$ 38.7 $&$  -  $&$ 0.57 $&$  -  $&$ 1.9 $&$ 
     0.015 $&$ 19.5 $&$ 3.48\e{1} $ \\ 
K2-137 b &$  0.46 $&$  0.44 $&$ 99.1 $&$ <160 $&$ 0.89 $&$ <1300 $&$ 0.18 $&$ 
     0.006 $&$ 2.83 $&$ 3.70\e{1} $ \\ 
Kepler-445 b &$  0.18 $&$  0.21 $&$ 90.0 $&$  -  $&$ 1.6 $&$  -  $&$ 3.0 $&$ 
     0.023 $&$ 23.4 $&$ 3.97\e{1} $ \\ 
K2-151 b &$  0.47 $&$  0.45 $&$ 69.6 $&$  -  $&$ 1.5 $&$  -  $&$ 3.8 $&$ 
     0.037 $&$ 17.8 $&$ 4.63\e{1} $ \\ 
K2-239 b &$  0.40 $&$  0.36 $&$ 49.0 $&$  -  $&$ 1.1 $&$  -  $&$ 5.2 $&$ 
     0.044 $&$ 26.3 $&$ 5.39\e{1} $ \\ 
K2-242 b &$  0.38 $&$  0.37 $&$ 110 $&$  -  $&$ 2.5 $&$  -  $&$ 6.5 $&$ 
     0.049 $&$ 28.7 $&$ 5.88\e{1} $ \\ 
Kepler-446 b &$  0.22 $&$  0.24 $&$ 120 $&$  -  $&$ 1.5 $&$  -  $&$ 1.6 $&$ 
     0.016 $&$ 14.3 $&$ 8.16\e{1} $ \\ 
K2-124 b &$  0.39 $&$  0.39 $&$ 140 $&$  -  $&$ 2.9 $&$  -  $&$ 6.4 $&$ 
     0.049 $&$ 27.0 $&$ 9.19\e{1} $ \\ 
K2-9 b &$  0.30 $&$  0.31 $&$ 83.2 $&$  -  $&$ 2.3 $&$  -  $&$ 18 $&$      0.091
 $&$ 63.1 $&$ 1.08\e{2} $ \\ 
K2-45 b &$  0.50 $&$  0.45 $&$ 502 $&$  -  $&$ 6.7 $&$  -  $&$ 1.7 $&$ 
     0.022 $&$ 10.7 $&$ 1.10\e{2} $ \\ 
Kepler-504 b &$  0.33 $&$  0.33 $&$ 75.0 $&$  -  $&$ 1.6 $&$  -  $&$ 9.5 $&$ 
     0.061 $&$ 39.6 $&$ 1.28\e{2} $ \\ 
K2-239 d &$  0.40 $&$  0.36 $&$ 49.0 $&$  -  $&$ 1.1 $&$  -  $&$ 10. $&$ 
     0.069 $&$ 40.9 $&$ 1.30\e{2} $ \\ 
K2-239 c &$  0.40 $&$  0.36 $&$ 49.0 $&$  -  $&$ 1.00 $&$  -  $&$ 7.8 $&$ 
     0.058 $&$ 34.4 $&$ 1.35\e{2} $ \\ 
Kepler-1646 b &$  0.24 $&$  0.26 $&$ 81.0 $&$  -  $&$ 1.2 $&$  -  $&$ 4.5 $&$ 
     0.033 $&$ 27.3 $&$ 1.43\e{2} $ \\ 
K2-95 b &$  0.44 $&$  0.42 $&$ 181 $&$  -  $&$ 3.9 $&$  -  $&$ 10. $&$ 
     0.070 $&$ 35.6 $&$ 1.59\e{2} $ \\ 
K2-72 b &$  0.27 $&$  0.33 $&$ 66.6 $&$  -  $&$ 1.1 $&$  -  $&$ 5.6 $&$ 
     0.040 $&$ 26.1 $&$ 1.64\e{2} $ \\ 
K2-288 B b &$  0.33 $&$  0.32 $&$ 65.7 $&$  -  $&$ 1.9 $&$  -  $&$ 31 $&$ 
     0.164 $&$ 110 $&$ 2.66\e{2} $ \\ 
Kepler-732 c &$  0.49 $&$  0.46 $&$ 150 $&$  -  $&$ 1.3 $&$  -  $&$ 0.89 $&$ 
     0.014 $&$ 6.69 $&$ 2.82\e{2} $ \\ 
Kepler-1582 b &$  0.28 $&$  0.30 $&$ 112 $&$  -  $&$ 1.5 $&$  -  $&$ 4.8 $&$ 
     0.037 $&$ 26.2 $&$ 2.99\e{2} $ \\ 
K2-104 b &$  0.43 $&$  0.49 $&$ 190 $&$  -  $&$ 2.0 $&$  -  $&$ 2.0 $&$ 
     0.023 $&$ 10.1 $&$ 3.27\e{2} $ \\ 
K2-72 d &$  0.27 $&$  0.33 $&$ 66.6 $&$  -  $&$ 1.0 $&$  -  $&$ 7.8 $&$ 
     0.050 $&$ 32.6 $&$ 3.31\e{2} $ \\ 
K2-150 b &$  0.46 $&$  0.44 $&$ 110 $&$  -  $&$ 2.0 $&$  -  $&$ 11 $&$ 
     0.073 $&$ 35.5 $&$ 3.42\e{2} $ \\ 
Kepler-445 d &$  0.18 $&$  0.21 $&$ 90.0 $&$  -  $&$ 1.2 $&$  -  $&$ 8.2 $&$ 
     0.045 $&$ 45.8 $&$ 3.98\e{2} $ \\ 
Kepler-560 b &$  0.34 $&$  0.33 $&$ 88.0 $&$  -  $&$ 1.7 $&$  -  $&$ 18 $&$ 
     0.095 $&$ 62.2 $&$ 4.42\e{2} $ \\ 
K2-72 c &$  0.27 $&$  0.33 $&$ 66.6 $&$  -  $&$ 1.2 $&$  -  $&$ 15 $&$ 
     0.078 $&$ 50.8 $&$ 4.70\e{2} $ \\ 
K2-88 b &$  0.26 $&$  0.26 $&$ 111 $&$  -  $&$ 1.2 $&$  -  $&$ 4.6 $&$ 
     0.035 $&$ 28.5 $&$ 5.45\e{2} $ \\ 
Kepler-446 d &$  0.22 $&$  0.24 $&$ 120 $&$  -  $&$ 1.3 $&$  -  $&$ 5.1 $&$ 
     0.035 $&$ 31.6 $&$ 5.50\e{2} $ \\ 
Kepler-1650 b &$  0.33 $&$  0.33 $&$ 121 $&$  -  $&$ 0.96 $&$  -  $&$ 1.5 $&$ 
     0.018 $&$ 11.7 $&$ 5.55\e{2} $ \\ 
K2-72 e &$  0.27 $&$  0.33 $&$ 66.6 $&$  -  $&$ 1.3 $&$  -  $&$ 24 $&$ 
     0.106 $&$ 69.0 $&$ 5.59\e{2} $ \\ 
Kepler-446 c &$  0.22 $&$  0.24 $&$ 120 $&$  -  $&$ 1.1 $&$  -  $&$ 3.0 $&$ 
     0.025 $&$ 22.2 $&$ 5.64\e{2} $ \\ 
K2-71 b &$  0.48 $&$  0.43 $&$ 154 $&$  -  $&$ 2.2 $&$  -  $&$ 7.0 $&$ 
     0.056 $&$ 27.9 $&$ 5.71\e{2} $ \\ 
K2-89 b &$  0.35 $&$  0.32 $&$ 86.2 $&$  -  $&$ 0.62 $&$  -  $&$ 1.1 $&$ 
     0.015 $&$ 9.82 $&$ 5.73\e{2} $ \\ 
K2-83 b &$  0.48 $&$  0.42 $&$ 126 $&$  -  $&$ 1.2 $&$  -  $&$ 2.7 $&$ 
     0.030 $&$ 15.4 $&$ 6.68\e{2} $ \\ 
Kepler-732 b &$  0.49 $&$  0.46 $&$ 150 $&$  -  $&$ 2.2 $&$  -  $&$ 9.5 $&$ 
     0.069 $&$ 32.3 $&$ 7.55\e{2} $ \\ 
K2-264 b &$  0.50 $&$  0.47 $&$ 187 $&$  -  $&$ 2.2 $&$  -  $&$ 5.8 $&$ 
     0.050 $&$ 23.0 $&$ 8.72\e{2} $ \\ 
K2-14 b &$  0.47 $&$  0.45 $&$ 368 $&$  -  $&$ 4.8 $&$  -  $&$ 8.4 $&$ 
     0.063 $&$ 29.9 $&$ 9.40\e{2} $ \\ 
Kepler-1649 b &$  0.22 $&$  0.25 $&$ 92.4 $&$  -  $&$ 1.1 $&$  -  $&$ 8.7 $&$ 
     0.051 $&$ 44.2 $&$ 1.01\e{3} $ \\ 
K2-125 b &$  0.49 $&$  0.40 $&$ 125 $&$  -  $&$ 2.2 $&$  -  $&$ 22 $&$ 
     0.121 $&$ 65.0 $&$ 1.10\e{3} $ \\ 
Kepler-1308 b &$  0.35 $&$  0.34 $&$ 73.0 $&$  -  $&$ 0.52 $&$  -  $&$ 2.1 $&$ 
     0.023 $&$ 14.3 $&$ 1.44\e{3} $ \\ 
K2-83 c &$  0.48 $&$  0.42 $&$ 126 $&$  -  $&$ 1.5 $&$  -  $&$ 10.0 $&$ 
     0.071 $&$ 36.5 $&$ 1.87\e{3} $ \\ 
Kepler-1124 b &$  0.35 $&$  0.34 $&$ 177 $&$  -  $&$ 1.3 $&$  -  $&$ 2.9 $&$ 
     0.028 $&$ 17.5 $&$ 1.98\e{3} $ \\ 
K2-264 c &$  0.50 $&$  0.47 $&$ 187 $&$  -  $&$ 2.7 $&$  -  $&$ 20 $&$ 
     0.113 $&$ 51.6 $&$ 2.15\e{3} $ \\ 
Kepler-296 c &$  0.50 $&$  0.48 $&$ 226 $&$  -  $&$ 2.0 $&$  -  $&$ 5.8 $&$ 
     0.052 $&$ 23.3 $&$ 3.13\e{3} $ \\ 
Ross 458 c &$  0.49 $&$  0.63 $&$ 11.5 $&$ 2000 $&$ 14 $&$ 4.3 $&$ 21000000 $&$ 
  1168.000 $&$ 399000 $&$ 4.79\e{3} $ \\ 
2MASS J02192210-3925225 b &$  0.11 $&$  0.28 $&$ 39.4 $&$ 4400 $&$ 16 $&$ 5.8
 $&$ 2100000 $&$    156.000 $&$ 120000 $&$ 6.04\e{3} $ \\ 
Kepler-1439 b &$  0.46 $&$  0.43 $&$ 215 $&$  -  $&$ 1.5 $&$  -  $&$ 8.1 $&$ 
     0.061 $&$ 30.4 $&$ 1.23\e{4} $ \\ 
Kepler-296 d &$  0.50 $&$  0.48 $&$ 226 $&$  -  $&$ 2.1 $&$  -  $&$ 20 $&$ 
     0.118 $&$ 52.8 $&$ 1.34\e{4} $ \\ 
Kepler-296 b &$  0.50 $&$  0.48 $&$ 226 $&$  -  $&$ 1.6 $&$  -  $&$ 11 $&$ 
     0.079 $&$ 35.4 $&$ 1.68\e{4} $ \\ 
K2-54 b &$  0.42 $&$  0.38 $&$ 176 $&$  -  $&$ 1.2 $&$  -  $&$ 9.8 $&$ 
     0.067 $&$ 38.0 $&$ 1.69\e{4} $ \\ 
Kepler-779 b &$  0.46 $&$  0.44 $&$ 224 $&$  -  $&$ 0.92 $&$  -  $&$ 7.1 $&$ 
     0.056 $&$ 27.3 $&$ 7.68\e{4} $ \\ 
Kepler-296 e &$  0.50 $&$  0.48 $&$ 226 $&$  -  $&$ 1.5 $&$  -  $&$ 34 $&$ 
     0.169 $&$ 75.7 $&$ 9.65\e{4} $ \\ 
Kepler-296 f &$  0.50 $&$  0.48 $&$ 226 $&$  -  $&$ 1.8 $&$  -  $&$ 63 $&$ 
     0.255 $&$ 114 $&$ 1.12\e{5} $ \\ 
Kepler-1652 b &$  0.40 $&$  0.38 $&$ 252 $&$  -  $&$ 1.6 $&$  -  $&$ 38 $&$ 
     0.165 $&$ 93.6 $&$ 1.17\e{5} $ \\ 
KOI-55 c &$  0.50 $&$  0.20 $&$ 1180 $&$ 0.67 $&$ 0.86 $&$ 5.7 $&$ 0.34 $&$ 
     0.008 $&$ 8.17 $&$ 1.41\e{6} $ \\ 
KOI-55 b &$  0.50 $&$  0.20 $&$ 1180 $&$ 0.45 $&$ 0.76 $&$ 5.5 $&$ 0.24 $&$ 
     0.006 $&$ 6.45 $&$ 1.45\e{6} $ \\ 

 \hline
\end{tabular}
\end{center}
\end{table*}

 We note the distinction between the radius of the planet and the radius of its dynamo region. As we show in Section \ref{sec.model}, the emitted radio power depends on the size of the planet and on its surface magnetic field strength, which depends upon how deep the dynamo region is within the planet. While the former can be derived from observations (item iii above), the latter depends on models. Recent work has suggested that rocky planets\footnote{In our sample,  average planetary densities $\langle \rho_p \rangle$ range between 1.5 and 7.5 g cm$^{-3}$, with the exception of K2-137 b (Table \ref{table_physics}).}  have magnetic dipole moments similar to that of the Earth or smaller \citep[][]{2019MNRAS.485.3999M}. However, depending on its composition, it is possible instead that terrestrial planets might host a stronger magnetic field than that of the Earth -- this has been suggested for K2-229b (not in our sample), given its Mercury-like composition and large radius \citep{2018arXiv181109198A}. Due to our lack of knowledge in planetary field strengths, in this paper, we assume three different planetary magnetic fields: 0.1~G, 1~G and 10~G.

%
\section{Model of planetary radio emission}\label{sec.model}
When the stellar wind interacts with a planet's magnetic field, it flows around the planet, confining the planet's magnetic field to a cavity surrounding the planet \citep{1930Natur.126..129C}. The extension of this cavity, known as the magnetosphere, can be calculated using pressure balance between the total pressure of the stellar wind ($p_{\rm sw}$) and the planet's  magnetic pressure ($p_B$). The total pressure of a stellar wind is the sum of its ram, thermal, and magnetic pressures. Only the wind's ram pressure is taken into account in this work, thus $p_{\rm sw}= \rho_{\rm sw} u^2$, where $\rho_{\rm sw} $ is the  mass density of the stellar wind at the orbit of the planet and $\vec{u} = \vec{u}_{\rm sw} - \vec{v}_K $ is the  velocity of the stellar wind particles in the reference frame of the planet, with $u_{\rm sw}$ being the velocity of the  stellar wind at the orbit of the planet  and $v_K$ the  Keplerian velocity of the planet (we assume circular orbit for simplicity, as 84\% of the  planets studied here have orbital eccentricity <0.2). The pressure due to the planet's intrinsic magnetic field is $p_B={B^2}/{8\pi}$, with $B$ being the planetary magnetic field intensity at a certain radius $r_M$ from the planet, where the interaction with the stellar wind happens. Assuming a dipolar planetary magnetic field, we have that $ B = B_p (R_p / r_M)^3$, where $B_p$ is the planet's surface magnetic field strength at the pole, and $R_p$ is the planet's radius. From this, we can estimate the magnetopause distance as
\begin{equation}\label{eq.rm}
\frac{r_M}{R_p} = \xi \left(\frac{(B_p/2)^2}{8\pi \rho_{\rm sw} u^2}\right)^\frac{1}{6},
\end{equation}
 where $\xi = 2^{1/3}$ is a numerical factor that corrects for the effects of electrical currents that flow along the magnetopause \citep{2004pssp.book.....C}. In cases where the stellar wind ram pressure is too strong for the planet magnetic field to sustain a magnetosphere, Equation (\ref{eq.rm})  gives $r_M<R_p$. In these cases, we set $r_M=R_p$.
 
 To predict the radio power arising from the interaction with a stellar wind, we use the radiometric Bode's law, i.e., we assume a linear relationship between the emitted radio power of a planet ($P_{\rm radio}$) and the kinetic power of the incident solar/stellar wind  ($P_{\rm sw}$): $P_{\rm radio}= \eta_K P_{\rm sw}$, where $\eta_K$ is the efficiency ratio, which has been determined empirically for solar system planets to be $10^{-5}$  \citep{2007P&SS...55..598Z}. The dissipated kinetic power of the stellar wind as it interacts with a planetary magnetosphere is approximated as the ram pressure of the stellar wind times the cross sectional area of the planetary magnetosphere, at a relative velocity $u$ \citep{2007P&SS...55..598Z, 2012MNRAS.423.3285V} 
\begin{equation}
    P_{\rm sw} = \rho_{\rm sw} u^3 \pi r_M ^2 .
\end{equation}
Thus giving the planet's radio power as
\begin{equation}\label{eq.pradio}
    P_{\rm radio} = \eta_K \rho_{\rm sw} u^3\pi r_M ^2.
    \end{equation}
Although here we assume that the input power of planetary radio emissions comes from the dissipated kinetic energy of the stellar wind upon interacting with the planet's magnetosphere \citep[e.g.][]{2004ApJ...612..511L}, scalings with the Poynting flux of the stellar wind also exist \citep{2007P&SS...55..598Z} . As we do not have knowledge of the magnetic field strength of the stars in our sample, we opted to use the scaling with the kinetic power of the stellar wind. Radio emission can also be generated in the interaction with CMEs or in Jupiter-Io-like interactions ({\it unipolar interaction}, \citealt{2007A&A...475..359G}). While the former can increase several orders of magnitude the radio power, the latter would likely generate non-detectable emission \citep{2007A&A...475..359G}. We will come back to the interaction with CMEs in Section \ref{sec.app2}.

Planetary radio emissions are observed to originate from a planet's auroral regions, which are strongly magnetised annular rings around a planet's magnetic poles \citep{1975JGR....80.4675S}. The boundary of a planet's auroral ring is thought to occur approximately where a planet's closed and open magnetic field lines meet \citep{2001JGR...106.8101H}, and is also referred to as the planet's polar cap \citep{2010Sci...327.1238T,2013A&A...557A..67V}. Planetary aurorae  are powered by magnetospheric currents that cause the precipitation of energetic electrons in the polar regions of the upper atmosphere \citep{WuLee1979, Hallinan2015}, leading to radio emissions via the cyclotron-maser instability mechanism \citep{2007P&SS...55..598Z}. The electrons are theorised to be accelerated along high-latitude field lines due to energy released during magnetic field line reconnection events in the nightside of the planet. 
In this work, we assume that radio emissions originate from the boundary of open and closed magnetic field lines at the colatitude of the planet's polar cap, as in  \citet{2017A&A...602A..39V}. This colatitude ($\alpha_0$) is related to the planet's magnetospheric size  and radius  as \citep{2011AN....332.1055V}
\begin{equation}\label{eq.alpha}
    \alpha_0 = \arcsin\left[\left(\frac{R_p}{r_M}\right)^{1/2}\right]
\end{equation}
Assuming dipolar magnetic fields, the surface magnetic field strength at this colatitude is given as
\begin{equation}
    B(\alpha_0) = \frac{1}{2}B_p[1 + 3\cos^{2}(\alpha_0)]^{1/2}. 
\end{equation}

When assessing the detectability of exoplanetary radio emissions, one needs to consider their frequency of emission and flux density. Planetary radio emissions are produced at a frequency close to that of the local electron cyclotron frequency. We assume that the emission bandwidth $\Delta f$ is approximately equal to the  electron cyclotron frequency, thus
\begin{equation}
    \Omega_{\rm cyc} = \frac{e B}{2 \pi m_e c}, 
\end{equation}
where $e$ is the charge of the electron, $m_e$ is the electron mass, and $c$ is the speed of light. Therefore, planets will radiate at a radio frequency determined by the magnetic field strength within the region where electron cyclotron-maser instability can occur. The maximum frequency occurs at the point where the planetary magnetic field is maximum, i.e., near the polar cap regions at the planetary surface.  Thus, we have
\begin{equation}
    \Delta f = 2.8\left(\frac{B(\alpha_0)}{1{\rm G}} \right) {\rm MHz. }
\end{equation}

Finally, the planetary radio flux density, which is the strength of the signal that is detected on Earth, is
\begin{equation}\label{eq.Fluxradio}
    \phi_{\rm radio} = \frac{P_{\rm radio}}{d^{2}\omega \Delta f} ,
\end{equation}
where $d$ is the distance to the system, and $\omega$ is the solid angle of the hollow cone of emission \citep{2017A&A...602A..39V} 
\begin{equation}\label{eq.omega}
\omega= 2 \int_{\alpha_0-\delta\alpha/2}^{\alpha_0+\delta\alpha/2} \sin \alpha {\rm d} \alpha {\rm d} \varphi = 4 \pi [\cos(\alpha_0-\frac{\delta\alpha}{2}) - \cos (\alpha_0+\frac{\delta\alpha}{2})] ,
\end{equation}
where the thickness of the cone is  $\delta\alpha=17.5^{\rm o}$  \citep{2004JGRA..10909S15Z}. The factor of two in the previous equation accounts for emission coming from both northern and southern hemispheres.

\section{Estimating radio emission from M-dwarf planets}\label{sec.results}
In order to calculate the stellar wind power dissipated in the planet's magnetosphere, we need values of the local density $\rho_{\rm sw}$ and velocity $u_{\rm sw}$ of the stellar wind. Since stellar winds of M dwarfs are poorly constrained, we proceed with two different approaches.

\subsection{Approach 1: Using hydrodynamic models of stellar winds}\label{sec.app1}
In this first  approach, we model the stellar wind as being thermally driven, spherically symmetric and in steady state  \citep{1958ApJ...128..664P}. For that, we solve the momentum equation of the stellar wind
\begin{equation}\label{eq.momentum}
\rho u \frac{du}{dR} = -\frac{G M_\star}{R^2} -\frac{dp}{dR}
\end{equation}
where $p = \rho k_B T/\bar{m}$ is the stellar wind thermal pressure, with $k_B$ being the Boltzmann constant, $T$ the isothermal wind temperature and $\bar{m}=0.5 m_p$ is the average mass of the stellar wind particle, taken to be a fully ionised, plasma of hydrogen (whose mass is $m_p$). The radial distance from the star is taken to be $R$ in Equation (\ref{eq.momentum}). Once a solution for the wind velocity $u(R)$ is found, the density profile is found from mass conservation $\rho(R) u(R) R^2 = {\rm constant}$. The local conditions for the stellar wind is found then at the semi-major axis of the planet $R=a$, and thus $\rho_{\rm sw} \equiv \rho(a) $ and $u_{\rm sw} \equiv u(a) $. 

The stellar masses $M_\star$ and radii $R_\star$  were taken from the NASA Exoplanet Archive (see Section \ref{sec.candidate}). An uncertainty in the stellar mass affects the stellar wind profile through the gravity force (second term in equation \ref{eq.momentum}). A variation of  $\pm 10\%$ in mass for a 0.5-$M_\odot$ star causes a negligible change in the terminal velocity (within $ \mp 1$\%). The density profile is more affected though, leading to changes in mass-loss rates of a factor of $0.6$ and $1.6$ for variations of $-10\%$ and $+10\%$ in stellar masses, respectively. However, values of the temperature and base density of  stellar winds are the most uncertain in our models. We adopt three different sets of temperature and base density. In our fiducial model (Section \ref{sec.fiducial}), for simplicity, we assume solar values for all the stars in our sample, with a temperature of  $1.56\times 10^6$~K, and base number density (electrons and protons) of $2.2 \times 10^{8}$~cm$^{-3}$ \citep{2018MNRAS.476.2465O}. For the second and third models, we vary the temperature and density by factors of 2 (Section \ref{sec.wind_disc}).

\subsubsection{Fiducial wind model (solar-like wind)}\label{sec.fiducial}
\begin{figure*}
\begin{center}
 	\includegraphics[width=.47\textwidth]{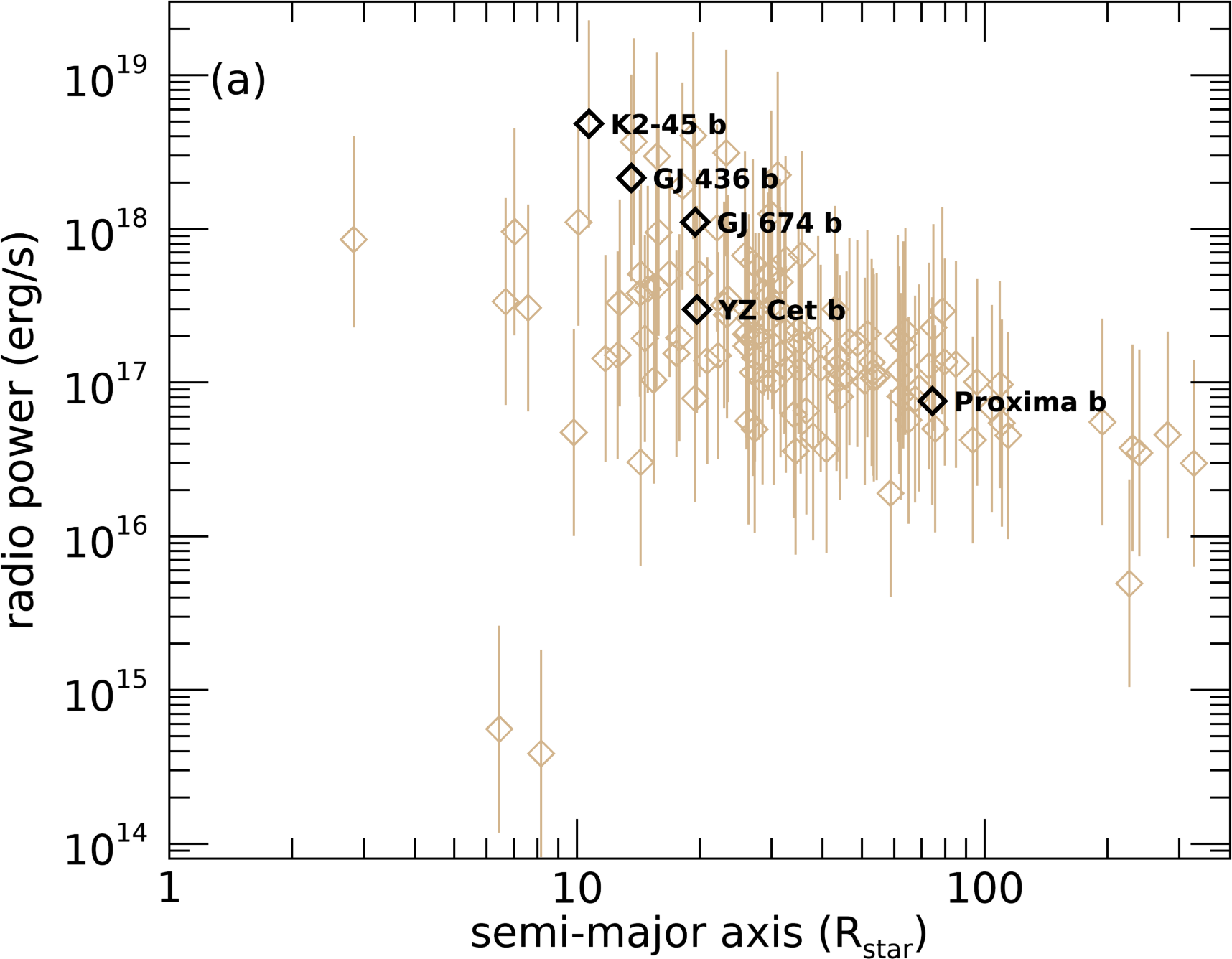}
 	\includegraphics[width=.47\textwidth]{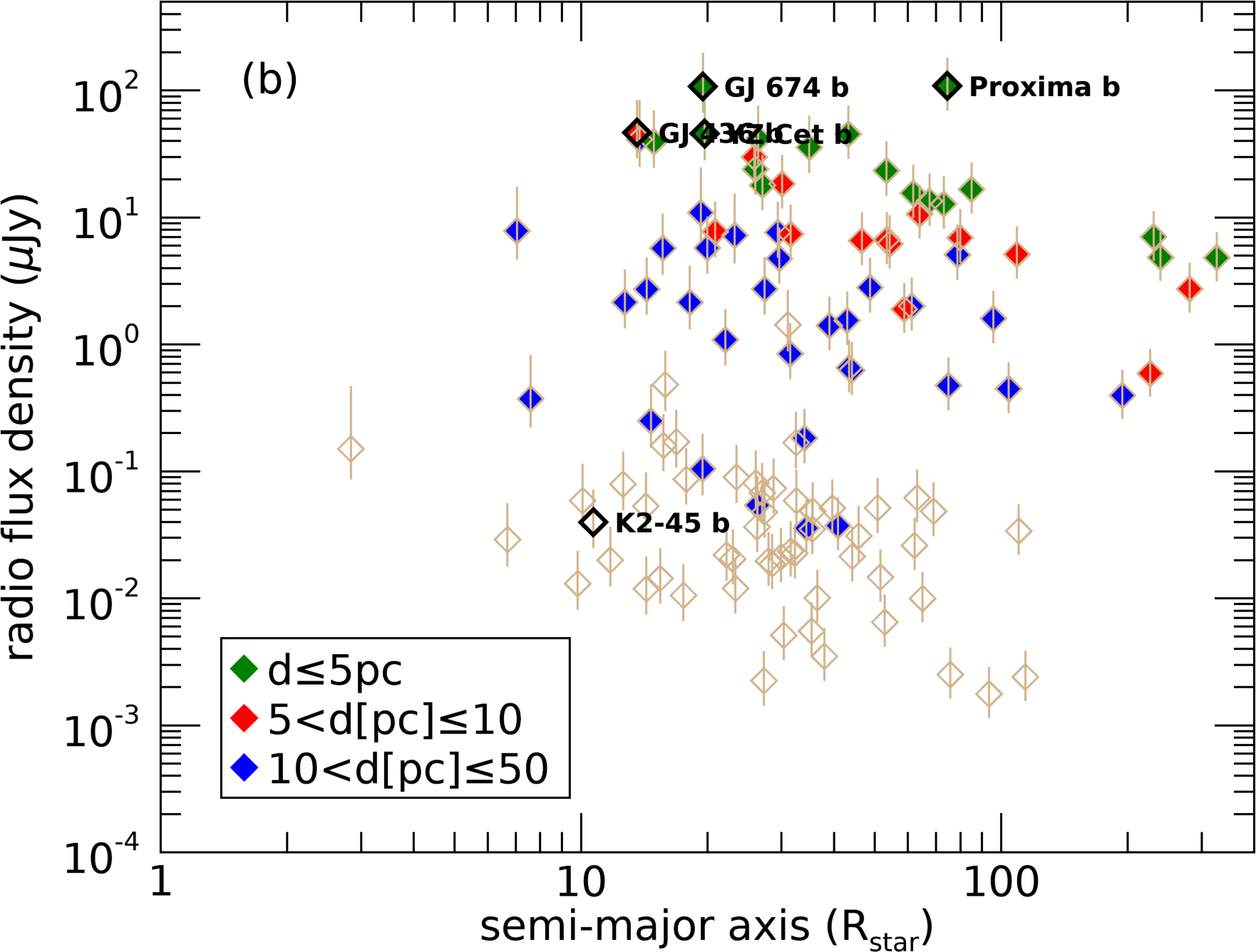}
  \caption{Radio power (a) and radio flux densities (b) of exoplanets listed in Table \ref{table_physics}. 
 The symbols represent a magnetic field strength of 1G, with the bars representing calculations done for 0.1G and 1G. The corresponding frequencies of emission are 2.5, 0.22 and 27~MHz, respectively. The values shown here are very sensitive to assumptions of stellar wind models. A model that is independent of the stellar wind driving mechanism is presented in Section \ref{sec.app2}. Distance is depicted as symbol colour in panel b.}
  \label{fig.radio}
  \end{center}
\end{figure*}

In this subsection, we calculate the planetary radio emission using the fiducial stellar wind model. 
Our computed radio power for each planet are shown in Figure \ref{fig.radio}a, for 1~G (symbols),  0.1~G (lower values of bars), and 10~G (upper values of bars). Our plot shows $P_{\rm radio}$ as a function of semi-major axis, where we see a trend of decreasing  $P_{\rm radio}$ with increasing semi-major axis (we normalise the semi-major axis with stellar radius, as this is the relevant distances in stellar wind computations, \citealt{2013A&A...557A..67V}). The spread seen in this correlation is due to the fact that planets are orbiting different stars of varying mass and radius, resulting in different local stellar wind properties, and that each planet has a different size. {K2-45b} has the strongest radio power of the sample ($10^{18}$~erg/s, for $B_p=0.1$~G), mostly due to its small semi-major axis (0.022 au $\simeq 11 R_\star$) -- planetary radio power  is strongest for close-in planets, where the stellar wind ram pressure is higher.

Figure \ref{fig.radio}b shows our calculated radio flux density  for the exoplanets in our sample. Again, these results are  for three magnetic field strengths: 1~G (symbols),  0.1~G (upper values of bars), and 10~G (lower values of bars). For these magnetic field strengths, emissions would occur at a maximum frequency of approximately 2.5, 0.22 and 27~MHz, respectively, averaged among the planets in our sample. Though K2-45b has the highest radio power of all of the planets, it has a very low flux density of 0.07~$\mu$Jy, for a 0.1G field. GJ 674 b and Proxima b have the highest flux density of 200 and 180~$\mu$Jy, respectively, followed by YZ Cet b, GJ 1214 b, GJ 436 b with flux densities around 83~$\mu$Jy, for a 0.1G field. K2-45b and some of these planets are highlighted in Figure \ref{fig.radio}. From the right panel, there is no clear correlation (i.e., large spread) between radio flux density and semi-major axis, that was seen in the plot of radio power (left panel). This increased spread is due to the variety of distances to the systems (depicted in the colours of the symbols). 
An increased magnetic field strength results in a small decrease in flux density, in spite of an increase seen in $P_{\rm radio}$.  For example, in the case of Proxima b, increasing the dipolar field from $1~$G to $10~$G, results in a flux that is a  factor of 1.5 smaller.  In our sample, a 2-order-of-magnitude increase in $B_p$ from 0.1~G to 10~G causes a decrease in radio fluxes by only a factor of 2.2 to 5.2, with an average of $\sim 2.7$. For the 1-G case,  solid angles  range from 0.4 to 2.8~sr, with a peak distribution at 1.7~sr,  similar to the Jovian value of $\sim$1.6~sr \citep{2004JGRA..10909S15Z}.

The symbol colours shown in Figure \ref{fig.radio}b depict distance to the system, with green representing objects that are closer to us, within 5 pc, and red representing objects at a distance between 5 and 10 pc, and blue between 10 and 50 pc. For reference, the expected value of Jupiter's emission, if Jupiter were at a distance of 50 pc, is at around $1~\mu$Jy \citep{2000ApJ...545.1058B}. For $B_p=0.1$~G, the flux densities range from $10^{-6}$ to 200 $\mu$Jy.  There is a tight negative correlation between flux and distance due to the inverse square relationship between the two (Equation~\ref{eq.Fluxradio}), which explains why the green and red symbols are all at the upper part of Figure \ref{fig.radio}b, with the more distant systems (empty symbols) at the bottom part. The top-4 exoplanets with the strongest flux densities (Proxima b, GJ 674 b, GJ 436 b, YZ Cet b) orbit some of the closest known M-dwarf planetary systems (1.3, 4.55, 9.76 and 3.6pc, respectively).  In contrast, {K2-45b}, which has the highest radio power, is more than 500 pc away, thus explaining its low flux density.

The second largest radio flux density among the planets in our study is that of Proxima b.  Given that Proxima b is an Earth-sized planet, \citet{2016ZuluProxB} estimated its magnetic field to be in the range of 0.1 to 0.3 G. From our models with $B_p=0.1$~G, we found a flux density of 180~$\mu$Jy. Recently,  \citet{2017ApJ...849L..10B} predicted Proxima b can emit between a few mJy up to 1~Jy depending on the assumed magnetic field of the planet (from 1G to 0.007G, respectively, cf.~their Figure 3, lower panel). For a magnetic field of $0.1$G, they predicted emission on the range of 4 to 30 mJy or even one order of magnitude higher if the dissipated power comes from the Poynting flux of the stellar wind. Our value (180~$\mu$Jy) differs from the ones (4 to 30 mJy) derived by \citet{2017ApJ...849L..10B} by one to two orders of magnitude. We attribute this disagreement to differences in the models -- for example, the frequency bandwidth in our model depends on the local value of the planetary surface field at the polar cap, while in  \citet{2017ApJ...849L..10B}'s model, it depends on the polar strength of the dipole. However, we believe that the main reason for the disagreement in radio flux lies is in the local condition of the stellar wind, which they assume to have a velocity of 1600~km/s (nearly 3 times faster than ours) and a density of about 1000 protons/cm$^{-3}$ (twice larger than ours). We investigate the influence of the stellar wind properties on planetary radio emission next. 

\subsubsection{Influence of the stellar wind properties}\label{sec.wind_disc}
Unfortunately, the values of the local stellar wind density $\rho_{\rm sw}$ and velocity $u_{\rm sw}$  are difficult to derive from observations. This is because the hot winds of cool dwarf stars are difficult to detect \citep{2004LRSP....1....2W}. In particular for M dwarfs, there is currently only few observational constraints, such as for Proxima Cen \citep{2001ApJ...547L..49W}, EV Lac \citep{2005ApJ...628L.143W}, GJ 436 \citep{2017MNRAS.470.4026V}, V374 Peg and HK Aqr \citep{2019MNRAS.482.2853J} and, so far, considerably fewer models of winds of M dwarfs \citep{2011MNRAS.412..351V, 2014MNRAS.438.1162V, 2016ApJ...833L...4G, 2017MNRAS.470.4026V,  2019MNRAS.482.2853J} than those proposed for sun-like stars. 

To quantify how the stellar wind affects radio flux densities, we perform the radio calculations for two additional wind models. Our original model assumes a wind temperature of $T \equiv T_\odot =1.56 \times 10^6$~K, and base number density of $n  \equiv n_\odot = 2.2 \times 10^{8}$~cm$^{-3}$. These values are typical of the solar wind. A second wind model has a temperature of $3.12 \times 10^6$~K, twice  as large,  and the same base density of $ 2.2 \times 10^8$~cm$^{-3}$. Finally, a third model has the same temperature as the second one, but with  a  twice as large base density of $4.4 \times 10^8$~cm$^{-3}$. We refer to each of these models as $\{T_\odot,n_\odot\}$,  $\{2T_\odot,n_\odot\}$ and  $\{2T_\odot,2n_\odot\}$, respectively.

The velocity of an isothermal wind is rather sensitive to temperature, but is independent of the choice of base density. This means that the stellar wind velocity at the planetary orbit is higher in model $\{2T_\odot,n_\odot\}$ than in model $\{T_\odot,n_\odot\}$. 
Because the velocity profile changes (more accelerated in model $\{2T_\odot,n_\odot\}$), the density structure also changes, seeing a slower decay with distance in the hotter of the two models. This also translates to a higher mass loss rate for the hotter model.  
The radial velocity  profile of the stellar wind for models $\{2T_\odot,n_\odot\}$ and  $\{2T_\odot,2n_\odot\}$ are the same, but the local wind densities are a factor of 2 higher in the model $\{2T_\odot,2n_\odot\}$. 
In our sample, local velocities for models $\{2T_\odot,n_\odot\}$ and $\{2T_\odot,2n_\odot\}$ are on average 1.7 times larger than for model $\{T_\odot,n_\odot\}$, ranging between factors of 1.5 and 2.9. 
Local densities are a factor 32 larger, on average, for model  $\{2T_\odot,n_\odot\}$ compared to $\{T_\odot,n_\odot\}$ (the factor ranges from 2.8 to 1500).
Mass-loss rates are on average a factor of 78 larger for $\{2T_\odot,n_\odot\}$  than for $\{T_\odot,n_\odot\}$ (factors ranging from 4.4 up to 3700), and then again a factor of 2 larger for $\{2T_\odot,2n_\odot\}$ compared to $\{2T_\odot,n_\odot\}$.

Overall, the stellar wind model $\{2T_\odot,2n_\odot\}$ presents the highest ram pressure as both the local wind density $\rho_{\rm sw}$ and velocity  $u_{\rm sw}$ are the largest of all the three models. This has two counter effects in planetary radio emission. While the radio power depends on the ram pressure (Equation \ref{eq.pradio}), it also depends on the area of the planetary magnetosphere, which decreases for higher ram pressure (Equation \ref{eq.rm}). Mathematically, we have that $P_{\rm radio} \propto \rho_{\rm sw} u^3 r_M^2 \propto  \rho_{\rm sw} u^3 /(\rho_{\rm sw} u^2)^{1/3} \propto (\rho_{\rm sw} u^2)^{2/3} u$. Hence, we expect that the hotter (i.e., higher velocities $u_{\rm sw}$ and thus higher relative velocities $u = (u_{\rm sw}^2 + v_K^2)^{1/2}$) and the denser the wind is, the dissipated radio power  will be higher. The dependence of the radio flux with the density and velocity is slightly more complicated (Equation \ref{eq.Fluxradio}), as we also need to consider the other two factors in the denominator, $\omega$ and $\Delta f$, which depend on the colatitude of the polar cap $\alpha_0$. In the limiting case of a radio emission that is powered by the kinetic energy of the stellar wind, we can write that $\phi_{\rm radio} \propto \rho_{\rm sw}^{1/2} u^2 f(\alpha_0)$ \citep[Equation B.1 in][]{2017A&A...602A..39V}. The function $f(\alpha_0)$ ranges between $\sim$0 and 3.3, for any given value of  $\alpha_0$ between $\sim$0 (small polar cap) and 90 degrees (crushed magnetosphere in the surface of the planet), respectively. 
Thus, the denser and hotter the wind is, the smaller the magnetosphere and hence the larger the value of $f(\alpha_0)$. Altogether, this leads to larger values of $\phi_{\rm radio}$. 

\begin{figure}
\begin{center}
 	\includegraphics[width=.47\textwidth]{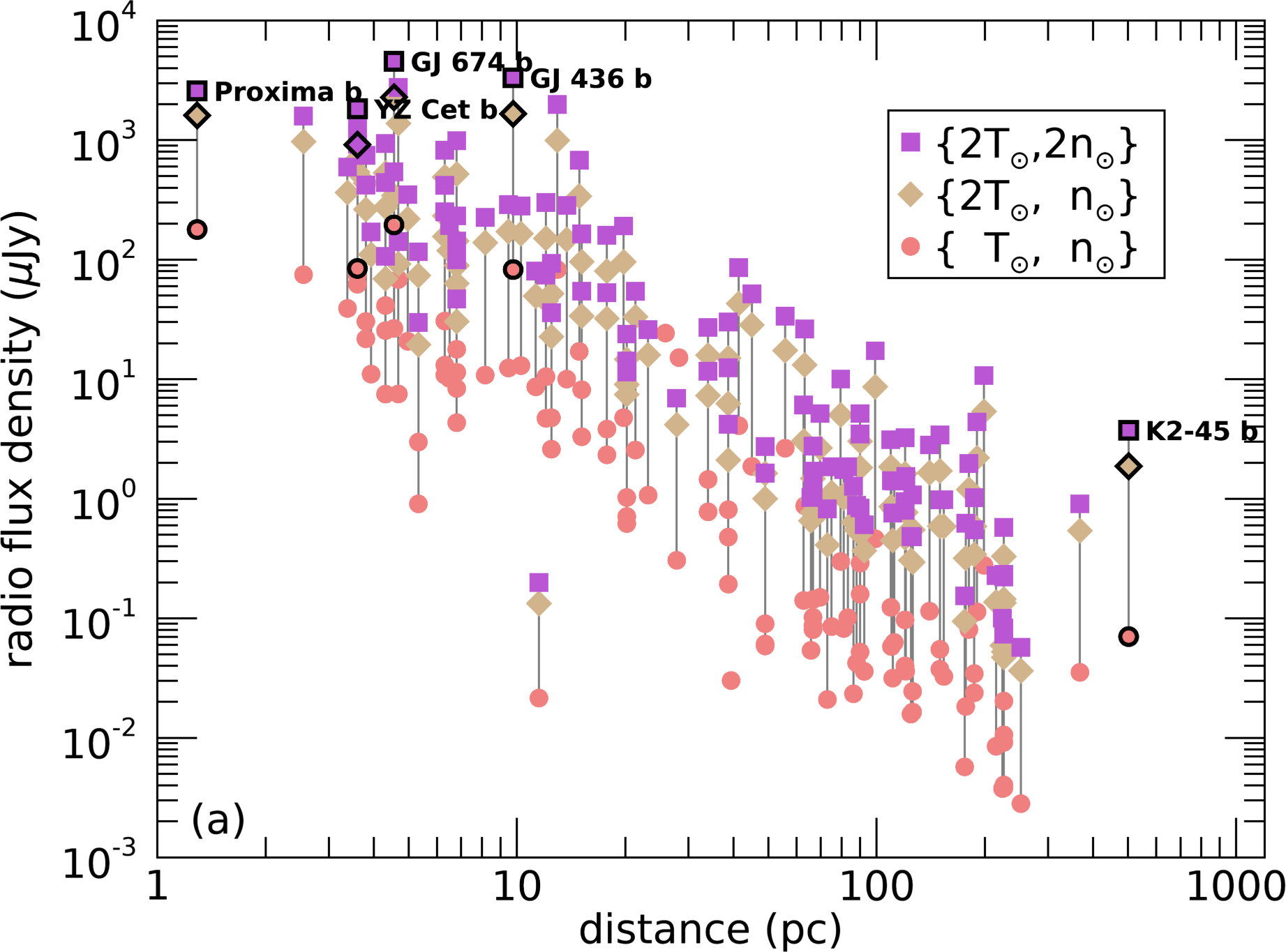}\\
 	\includegraphics[width=.47\textwidth]{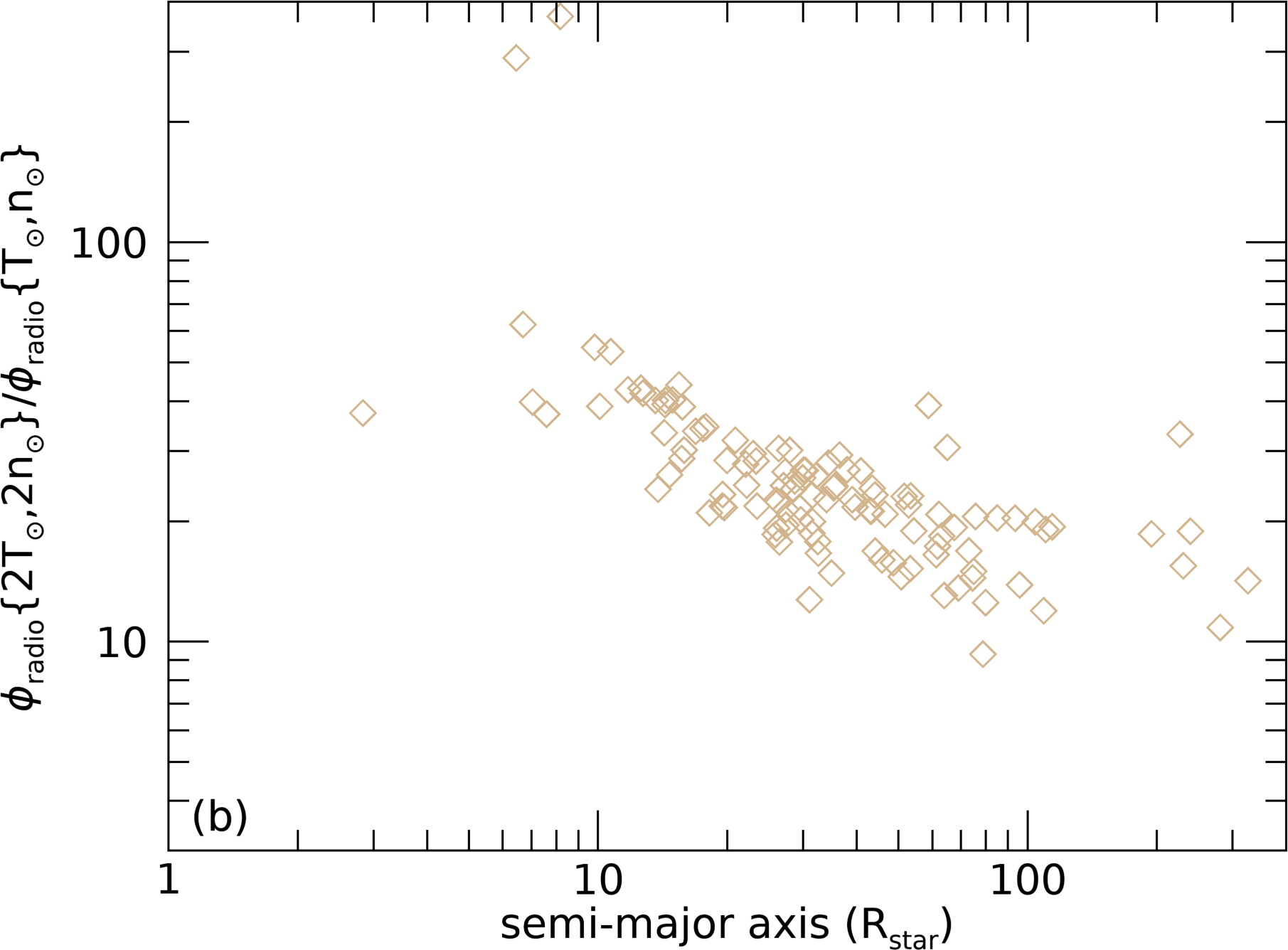}
  \caption{(a) Radio flux density predictions assuming a planetary dipolar field strength of 0.1 G and their dependence on distance to the system.  Three stellar wind conditions are investigated, ranging from typical base values of the solar wind and twice its temperature and density. Hotter and denser winds induce overall higher planetary radio emissions. (b) Ratio between the radio flux density derived with model  $\{2T_\odot,2n_\odot\}$ and $\{T_\odot,n_\odot\}$ as a function of orbital distance. Uncertainties in the stellar wind properties affect more the radio predictions of close-in planets. }
  \label{fig.wind_comparison}
  \end{center}
\end{figure}

This can indeed be seen in Figure \ref{fig.wind_comparison}a, where we present the radio flux densities of the exoplanets in our sample for the three different wind models. For this calculation, we keep the value of the planetary dipolar field strength fixed to $B_p=0.1$~G. The stellar wind model $\{2T_\odot,2n_\odot\}$ induces the highest radio emission for all planets, followed by model $\{2T_\odot,n_\odot\}$ and our typical solar wind value case (model $\{T_\odot,n_\odot\}$) leading to the smallest radio flux among all the three stellar wind models. We note the distance-squared decay of the radio flux very clearly in this figure -- the outlier point at 11.5pc, with low flux density, is Ross 458c, which has one of the largest semi-major axis in our sample (i.e., with weak stellar wind).
Figure \ref{fig.wind_comparison}b shows the ratio of radio fluxes predicted with model $\{2T_\odot,2n_\odot\}$ over those of model  $\{T_\odot,n_\odot\}$. Overall, the discrepancy between the two computed fluxes are factors between $\sim 10$ and 60, but reaches above 200 for two exoplanets seen in the figure. The discrepancy is larger for close-in planets, which is also the region favoured in the detection of exoplanets, due to detection bias.  

In the case of Proxima b, for example, the radio flux density increases from $0.18$mJy for model $\{T_\odot,n_\odot\}$ to $1.6$mJy for model $\{2T_\odot,n_\odot\}$ to $2.6$mJy for model $\{2T_\odot,2n_\odot\}$.Our model $\{2T_\odot,2n_\odot\}$ produces a  local wind velocity of 885~km/s  and proton density of about 5600 protons/cm$^{-3}$, which is approximately factors of 1/2 and 5, respectively, compared to the  values adopted in \citet{2017ApJ...849L..10B}. This explains why only model $\{2T_\odot,2n_\odot\}$ has a radio flux for Proxima b that is more similar to the ones derived in \citet{2017ApJ...849L..10B}. The most promising target, GJ 674 b, sees an increase in the radio flux density  from $0.2$mJy to $4.6$mJy by changing the density and temperature of the stellar wind model by factors of 2. 

\subsection{Approach 2: Deriving local properties of stellar winds that power detectable radio emission}\label{sec.app2}
As demonstrated in Figure \ref{fig.wind_comparison}, an uncertainty in the wind properties of M-dwarf stars can generate a significant uncertainty in the predicted radio emission. Which type of stellar wind would generate detectable radio emission?  Before we present the results of our computation, we stress that, contrary to the results presented in Section \ref{sec.app1}, the stellar winds we discuss in this section do not depend on any particular wind model. Rather, the results shown next only depend on the local property of the stellar wind at the orbit of planets and not on which physical mechanism is responsible for accelerating the winds to such velocities. 

 In the limiting case of a radio emission that is powered by the kinetic energy of the stellar wind, we have \citep[Equation B.1 in][]{2017A&A...602A..39V}
\begin{equation}\label{eq.phiradio3}
\phi_{\rm radio}  =  \eta_K^\prime  \frac{R_p^2}{d^2 }{\rho_{\rm sw}^{1/2} u^2} f(\alpha_0) \, ,
\end{equation}
where $\eta_K^\prime \simeq \eta_K 1.8\e{-8} = 1.8\e{-13}$ (in cgs units),  $u =(u_{\rm sw}^2 + v_K^2)^{1/2}$ is the relative velocity   and 
\begin{equation}\label{eq.falpha}
 f(\alpha_0)  =  \frac{\sin^2\alpha_0}{ [\cos(\alpha_0-\delta\alpha/2) - \cos (\alpha_0+\delta\alpha/2)] (1+3\cos^2 \alpha_0)^{1/2}} \, ,
\end{equation}
where we note that $0 \lesssim f(\alpha_0)< 3.3$. Note that our ignorance on $B_p$ is hidden in the function $ f(\alpha_0)$, through Equations (\ref{eq.rm}) and (\ref{eq.alpha}). Rewriting Equation (\ref{eq.phiradio3}), we have
\begin{equation}
{\rho_{\rm sw}^{1/2} u^2} = \frac{\phi_{\rm radio}  }{ \eta_K^\prime } \frac{d^2 }{R_p^2} \frac{1}{f(\alpha_0)}  \, .
\end{equation}
Let us assume now that  $ f(\alpha_0)$ takes its \textit{largest} value of $\simeq 3.3$. In this case, we estimate that the \textit{minimum} required property of the local stellar wind is
\begin{equation}\label{eq.flux_min}
\left\{ \left(\frac{n_{p, \rm sw}}{\rm {1\,cm}^{-3}}\right)^{1/2} \left(\frac{u}{1 \, \rm km/s}\right)^{2} \right\}_{\rm min} \simeq \frac{\phi_{\rm radio} / (1\,{\rm mJy)} }{ 7.6\e{11} } \left(\frac{d }{R_p}\right)^2 \, ,
\end{equation}
where we converted the total mass density in proton density ($ \rho_{\rm sw} = n_{p, \rm sw} m_p$).  Alternatively, we can rewrite Equation (\ref{eq.flux_min}) in terms of mass-loss rates $\dot{M}$
\begin{equation}\label{eq.flux_min2}
\left\{ \left(\frac{\dot{M}}{\msano}\right)^{1/2} \left(\frac{u}{1 \, \rm km/s}\right)^{3/2} \right\}_{\rm min} \simeq \frac{\phi_{\rm radio} / (1\,{\rm mJy)} }{ 2.8\e{20} } \left(\frac{d }{R_p}\right)^2  \frac{a}{\rm 1\, au}\, ,
\end{equation}
where we used $\dot{M} = \rho_{\rm sw} u_{\rm sw} 4 \pi a^2$, for a spherically symmetric stellar wind. 
Equations (\ref{eq.flux_min}) and (\ref{eq.flux_min2}) imply that, in case of radio detection, we cannot decouple the local density (or mass-loss rate) from the local velocity of the stellar wind, unless we provide additional hypotheses (or additional observations). 

\begin{figure}
\begin{center}
 	\includegraphics[width=.47\textwidth]{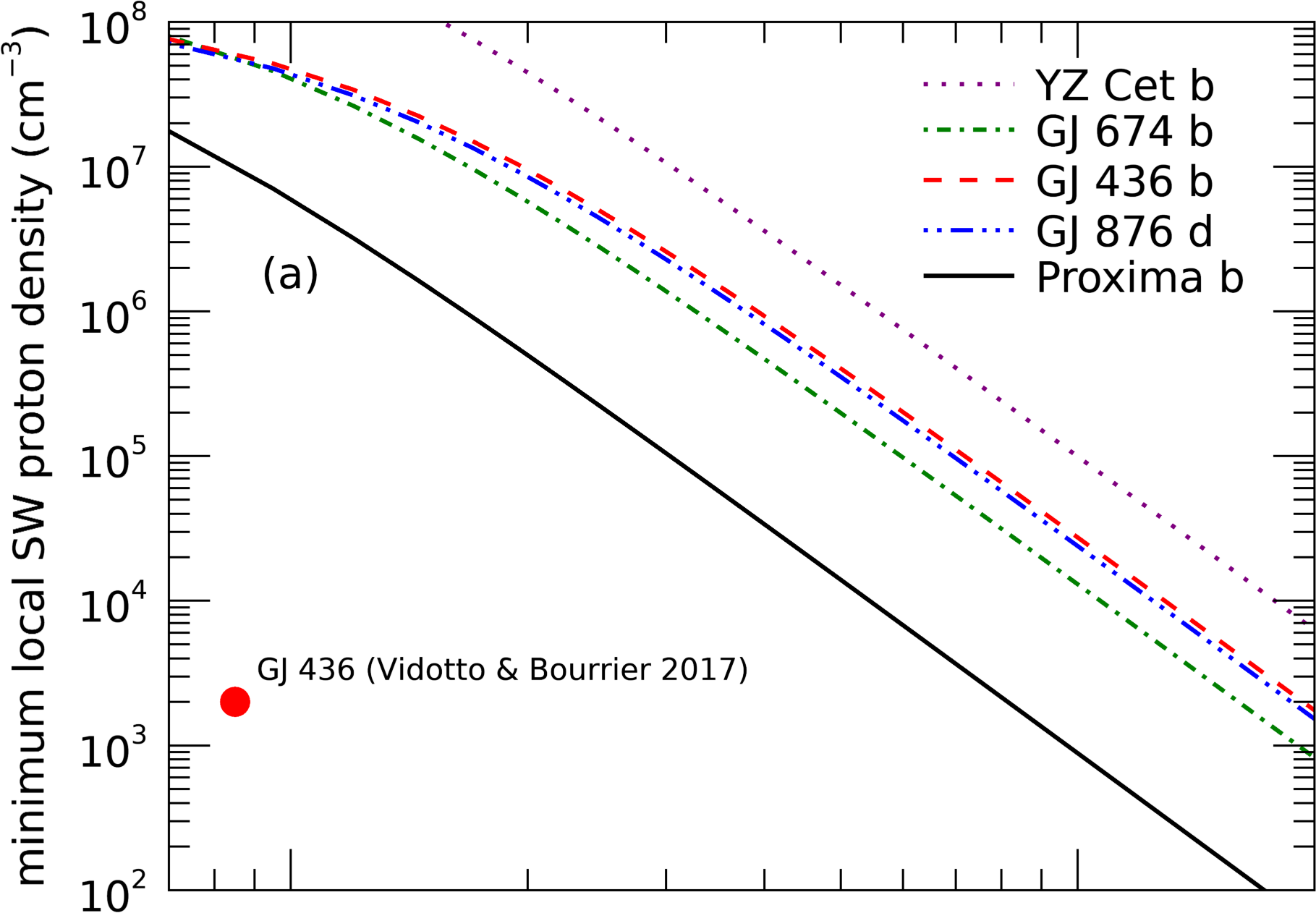}\\
 	\includegraphics[width=.47\textwidth]{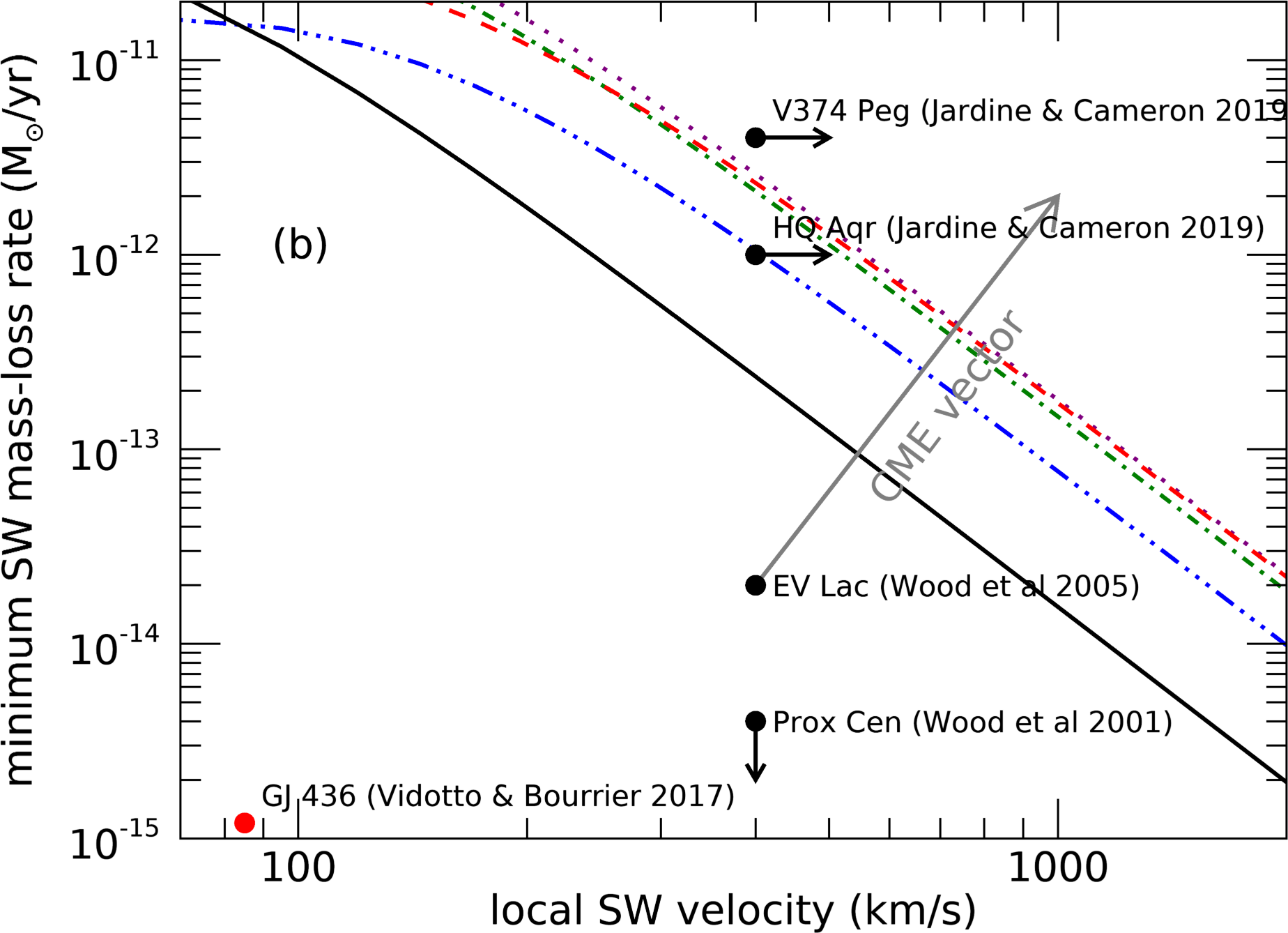}
  \caption{Minimum stellar wind properties required to generate a radio flux density of 1~mJy for selected exoplanets. (a) Minimum local density of the stellar wind and (b) Minimum mass-loss rate as a function of the local stellar wind velocity. Contrary to Figure \ref{fig.radio}, these plots do not depend on stellar wind models (i.e., their mechanism of acceleration), but rather only on the local conditions of the stellar wind. Some M dwarfs with observational constraints for their stellar winds are shown. Coronal mass ejections are expected to increase velocities and mass-loss rates, which would momentarily power radio emission to detectable levels.}
  \label{fig.min_wind}
  \end{center}
\end{figure}

The curves in Figure \ref{fig.min_wind} (Equations (\ref{eq.flux_min}) and (\ref{eq.flux_min2}) in the top and bottom panels, respectively) show the dependences of the stellar wind properties for a given radio flux density of 1mJy. We chose this (optimistic) value of radio flux as this would generate detectable radio emission (i.e., above instrument sensitivity limit) for present-day instruments (Section \ref{sec.discussion}). If the sensitivity limit becomes lower, then the curves in Figure \ref{fig.min_wind} would shift down, i.e., the minimum required wind properties become smaller. We only show the curves for some selected systems, for which the right-hand side of Equation (\ref{eq.flux_min}) provides the minimum values in our sample (i.e., systems that have the smallest values of $d/R_p$ in the sample). The inflexion seen is some of the curves towards low stellar wind velocities is because Equations (\ref{eq.flux_min}) and (\ref{eq.flux_min2}) are written with respect to relative velocity $u$ while the curves are plotted against $u_{\rm sw}$. The several points shown in Figure \ref{fig.min_wind} are constraints derived from observations and will be described in Section \ref{sec.const}.

Table \ref{table_physics} shows, for all the exoplanets in our sample,   the wind constant $\kappa_{\rm sw}$ that obeys the relation 
 \begin{equation}
\dot{M} \gtrsim \kappa_{\rm sw} /u_{\rm sw}^3,
 \end{equation}
with mass-loss rate $\dot{M}$ given in $\msano$ and local stellar wind velocity $u_{\rm sw}$ in km/s, where
 \begin{equation}
\kappa_{\rm sw} = \left[ \frac{\phi_{\rm radio} / (1\,{\rm mJy)} }{ 2.8\e{20} } \left(\frac{d }{R_p}\right)^2  \frac{a}{\rm 1\, au} \right]^2 \, 
 \end{equation}
 (cf.~Equation \ref{eq.flux_min2}). 
These equations provide the minimum mass-loss rates required for a stellar wind to power a given radio flux $\phi_{\rm radio} $. The best candidates for radio detection among the planets in our sample are Proxima b, GJ 436 b,  GJ 674 b, GJ 876 b and YZ Cet b. To be detected with current instrumentation, the quiescent winds of their hosts should have $\dot{M} u_{\rm sw}^{3} \gtrsim 1.8\e{-4}  \msano ({\rm km/s})^{3} $. As we will see next, the quiescent winds of some of these stars are likely too weak to power detectable radio emission.

\subsubsection{Observational constraints of M dwarf winds}\label{sec.const}
 In order to assess the observability of radio emission from M dwarf exoplanets we add some available observational constraints of the winds of M dwarfs to Figure  \ref{fig.min_wind} -- these are represented by filled circles.  Values of mass-loss rates and wind velocities\footnote{The quoted velocities are terminal stellar wind velocities, except for GJ436, which we use the local velocity from  \citet{2017MNRAS.470.4026V}. Note that the solar wind values at 1au match values derived for EV Lac  \citep{2005ApJ...628L.143W}.} are quoted for  Proxima Cen \citep{2001ApJ...547L..49W}, EV Lac \citep{2005ApJ...628L.143W}, GJ 436 \citep{2017MNRAS.470.4026V}, V374 Peg and HK Aqr \citep{2019MNRAS.482.2853J}. For the last two objects, we assume a (terminal) velocity of $\sim 400$~km/s but note that for active M dwarfs, these values are likely higher \citep{2011MNRAS.412..351V, 2018MNRAS.475L..25V}. Of these five objects, only two have detected exoplanets and we explore them in  more details next. 

Assuming a threshold of 1mJy, which is achievable with present-day instruments, we have that, for Proxima b to have a detectable radio emission, the stellar wind should have a minimum mass-loss rate of 
\begin{equation}
\dot{M} \gtrsim 1.6\e{-5} /u_{\rm sw}^3 \,\,\,\,\, {\rm for~Proxima~b.} 
\end{equation}
 Note that the typical values used in \citet{2017ApJ...849L..10B} for density (1000~cm$^{-3}$) and velocity (1600~km/s) are indeed above our minimum line for Proxima b (Figure \ref{fig.min_wind}a), confirming the need of relatively high densities and velocities for a detectable radio emission from Proxima b. \citet{2001ApJ...547L..49W} derived a mass-loss rate of $<2\e{-15}~\msano$ with a terminal velocity of 400~km/s for a quiescent wind of Proxima Cen. These values are below the minimum stellar wind properties, as can be seen in  Figure \ref{fig.min_wind}b, which implies that such a quiescent wind would not drive radio emission above 1mJy level. 
 
 In the case of GJ 436 b, the required host star mass-loss rate is 
 \begin{equation}
 \dot{M} \gtrsim 1.8\e{-4} /u_{\rm sw}^3 \,\,\,\,\, {\rm for~GJ~436~b} ,
 \end{equation}
 (we here assumed that $u \simeq u_{\rm sw} $). 
The red circles in Figure \ref{fig.min_wind}  show the mass-loss rate, local proton density and stellar wind velocity derived for GJ 436 \citep{2017MNRAS.470.4026V}. Given these values, it is unlikely that radio emission from GJ 436 b would be detectable with present-day instruments.

Both the quoted values of Proxima Cen and GJ 436 are representative of quiescent winds. However, the passing of a CME would increase the  density and  velocity of local stellar winds. In the Sun, CMEs can reach velocities above 1000 km/s \citep[e.g.][]{2003ApJ...588L..53G}. \citet{2015ApJ...809...79O} estimate that CMEs in EV Lac could change the mass-loss rate by 2 orders of magnitude the values derived from \citet{2005ApJ...628L.143W}. We add therefore a ``CME vector'' to Figure \ref{fig.min_wind}b to indicate the general trend one would expect to achieve with episodic mass ejections. This means that, although quiescent stellar winds might not drive detectable planetary radio emissions, the increase in velocities and densities of CMEs might be able to reach values above our minimum thresholds in Figure \ref{fig.min_wind}b.

\section{Discussion}\label{sec.discussion}
Detection  of exoplanetary radio emission could help derive stellar wind properties. In this case, the detection frequency would allow one to assess the planetary magnetic field. The radio flux density, on the other hand, would then allow one to derive the stellar wind conditions. Given that the radio flux density is rather sensitive to stellar wind conditions (Figure \ref{fig.wind_comparison}a), radio detection could help us pinpoint the most likely stellar wind properties (Figure \ref{fig.min_wind}).

\subsection{Ground-based detectability of predicted emissions}
One of the key factors in the detection of radio emission from exoplanets is the frequency of emission. Since emission happens at the cyclotron frequency, knowing the magnetic field strength of the planet is a key element to help guide searches. Our models predict  a maximum emission bandwidth of 2.7 MHz for $B_p=1$~G. However, the Earth's ionosphere reflects electromagnetic radiation of frequencies below 10 MHz, therefore ground-based radio telescopes can not detect signals below this frequency. When the planetary magnetic field strength is increased to 10 G, the maximum emission bandwidth increases to 27 MHz, which is above the Earth's ionospheric cutoff. There are a number of radio arrays operating in the low frequency range (below $\sim 100$ MHz), such as the upgraded Ukrainian T-shaped Radio telescope (UTR-2), the Low-Frequency Array (LOFAR), the Murchison Widefield Array (MWA) and the Owens Valley Long Wavelength Array (OVRO-LWA). Currently,  LOFAR offers the best frequency of operation-sensitivity combination: \citet{2019A&A...622A...5D} estimate a sensitivity of about 3mJy/beam at 54MHz, or $\sim 10$~mJy for a 3$\sigma$ detection. For a  1-h integration time at 20 to 40 MHz, \citet{2011RaSc...46.0F09G} project a more optimistic sensitivity of about 3 to 30 mJy. The square kilometre array (SKA) will outperform LOFAR in sensitivity, however it is set to operate at frequencies above 50 MHz \citep{SKA2009,Grie2017}. For a planet to emit at a frequency of over 50 MHz, it would require a minimum magnetic field strength of $\sim$ 18G. 

 Knowledge of the magnetic fields of exoplanets is currently based on theoretical estimations and extrapolations of solar system models \citep[e.g.,][]{2009Natur.457..167C, 2012Icar..217...88Z, 2012A&A...546A..19G, 2016ZuluProxB}. Although some works have suggested means to derive planetary field strengths \citep[e.g.][]{2010ApJ...722L.168V, 2014Sci...346..981K}, there is still no definitive detection of exoplanetary magnetic fields. Recent studies suggest, however, that rocky planets, such as most of the planets in our sample, would have magnetic dipole moments that are at the most similar to that of the Earth \citep[][but see also \citealt{2018arXiv181109198A} for terrestrial exoplanets with higher predicted magnetic fields]{2019MNRAS.485.3999M}.  Therefore, if M-dwarf planets have magnetic field strengths $\lesssim 4$G, as is likely the case of Proxima b \citep{2016ZuluProxB}, radio emissions are not detectable from ground-based radio telescopes. In recent years, there has been discussion of construction of a radio array on the moon \citep{Lazio2011, Zarka2012Lunar}. Although perhaps not suitable for exoplanet radio emission detection, the Chang'e 4 lander carried a radio astronomy payload, and the Chang'e 4 orbiter carries the Netherlands-Chinese Low-frequency Explorer payload, which may represent first steps towards future radio detection of exoplanets.  Similarly, the Sun Radio Interferometer Space Experiment (SunRISE) concept may also present a path forward in future low-frequency space-based radio observatories. Space-based radio instruments would have an enormous impact on the search for exoplanetary radio emissions, allowing signals from low frequency sources, such as exoplanets, to be detected and characterised. The results found in this study, among many others, provide additional motivation for the building of a lunar array.

\subsection{Time-dependence of radio emissions}\label{sec.cme}
The predictions made here do not take into account potential variability of radio emission.  A planet's magnetosphere, for example, would vary in size due to the changing properties of stellar winds.  Jovian emissions are sporadic at times with transient periods of high-intensity emissions, which correspond to solar flares or coronal mass ejections \citep{1999JGR...10414025F}. Proxima Centauri, for example, is expected to have a highly variable stellar wind with frequent flaring events \citep{2016ApJ...829L..31D}. This could mean that radio emissions from Proxima b could vary in intensity, with periods of extremely intense radio emissions that could lie within the ground-based detectable range \citep{2017ApJ...849L..10B}. Rotation of planets can also cause changes in planetary radio emission. For example,  for the giant planets of the solar system, the observed flux densities of emission are modulated at the planetary rotation period. This modulation is caused by an offset between their magnetic and rotation axes \citep{2000ApJ...545.1058B, 2007P&SS...55..598Z}.  It remains to be seen whether such offsets would exist in exoplanets, and/or if they are sufficiently large  to cause a detectable variation in the intensity of radio emissions \citep{2018A&A...616A.182V}. 

In addition to short-term variability of planetary radio emission, caused by short-term variability in stellar winds (e.g., due to passing of a CME, Figure \ref{fig.min_wind}b), longer term changes in radio emission are also expected \citep{2011MNRAS.414.1573V}. For example, on scales of days to years, as the planet moves along its orbit, the local properties of stellar winds change. In all the wind models we presented here, the winds are assumed to be spherically symmetric. However, the complex geometries seen in stellar magnetism imply that stellar winds are not homogeneous \citep{2013MNRAS.436.2179L,2015MNRAS.449.4117V, 2016A&A...594A..95A, 2017ApJ...849L..10B}. On scales of years to decades, we also observe changes in stellar magnetism associated to cycles \citep{2016A&A...594A..29B,2016MNRAS.459.4325M, 2016MNRAS.462.4442S}, which means that stellar winds would also evolve in these timescales \citep{2011ApJ...737...72P, 2016MNRAS.459.1907N,2019arXiv190309871F}. Finally, on evolutionary timescales of Gyr, winds are believed to become more rarefied as the star evolves in the main sequence \citep{2014A&A...570A..99S,2015A&A...577A..28J,2015A&A...577A..98G,2015ApJ...799L..23M, 2019MNRAS.483..873O}.

\section{Conclusions}\label{sec.conclusions}
This work investigated the detectability of radio emission for currently known exoplanets orbiting M dwarf stars. After filtering for planets with host stars in the mass range between 0.1 and 0.5 M\textsubscript{$\odot$}, our sample contained 120 exoplanets orbiting M dwarf stars (Table \ref{table_physics}). The radio emission calculation was based on the radiometric Bode's law -- an empirical law determined for the solar system, which has since been extrapolated to extrasolar systems. The radiometric Bode's law depends on the local properties of the host-star wind. In the case of M dwarfs, these properties are still very poorly constrained, both from modelling and observational aspects. We therefore tackled this with two approaches. 

In the first approach, we used the host-stars characteristics to construct simple stellar wind models, extracted the wind properties at the orbits of our planets and calculated the radio flux densities of the exoplanets in our sample. Most previous work in this area have focused on hot Jupiters orbiting  F, G and K type stars, therefore our study filled the gap by estimating auroral emission from planets orbiting M dwarfs. We assumed these planets have dipolar fields and adopted three dipolar field strengths: 0.1~G, 1~G and 10~G.  Given that the radio emission occurs at the cyclotron frequency,  emissions would occur at a maximum frequency of approximately 0.22, 2.5 and 27~MHz (averaged for all the planets in the sample), respectively. Recent works have suggested that rocky planets, as most of the planets in our sample, are more likely to present magnetic moment similar to or smaller than that of the Earth (the Earth's dipolar field strength at poles is 0.3~G), so the lower magnetic field values of 0.1 and 1~G are preferred. Nevertheless, we showed that there is a weak dependence of the planet's radio flux density on its magnetic field strength, with a factor of 100 increase in magnetic field strength (from 0.1 to 10~G) resulting in a decrease in flux density by a factor of $\sim 2.7$ only.

In our sample, K2-45b has the largest radio power due to its large size. However, its estimated flux density is quite small, $< 10~\mu$Jy, due to it being at a distance of more than 500~pc from us. GJ 674 b, Proxima b, GJ 1214 b, GJ 436 b and YZ Cet b have the strongest flux densities despite their relatively low radio power. Using our simple stellar wind models, we predict flux densities for these exoplanets that are above the mJy level. The higher fluxes are due to their proximity to us as they all orbit stars less than 10 pc from us. Exoplanets of close proximity should be therefore the strongest candidates for future searches of detectable planetary radio emissions. Proxima b is of particular interest due to its proximity (1.3 pc). Given its terrestrial size, dynamo models suggest a magnetic field strength $<1$~G. If this is indeed the case, its emission frequency would be below the Earth's ionospheric cutoff point of 10 MHz and would not be possible to detect Proxima b's radio emission with ground-based observations.  

We demonstrated that the uncertain stellar wind properties of M dwarf stars can generate a significant uncertainty in the predicted radio emission. Although this might seem disappointing at first, we argued that detection of radio emission from exoplanets could  help us pinpoint the most likely  wind properties of M dwarfs, which are currently very poorly constrained by observations. Therefore, in our second approach, instead of basing radio estimates on unknown properties of stellar winds, we investigated which physical characteristics of the host-star wind are required in order for radio emission to be detectable with present-day instruments at the 1mJy level. We  found that the minimum mass-loss rates obey the relation $\dot{M} \gtrsim \kappa_{\rm sw} /u_{\rm sw}^3$, with mass-loss rate $\dot{M}$ given in $\msano$ and local stellar wind velocity $u_{\rm sw}$ in km/s. The derived values of $\kappa_{\rm sw}$, calculated assuming a  sensitivity limit of 1mJy, are presented in Table \ref{table_physics}. For Proxima~b, we found that $\dot{M} \gtrsim 1.6\e{-5} /u_{\rm sw}^3$ and for GJ 436 b, $\dot{M} \gtrsim 1.8\e{-4} /u_{\rm sw}^3$. Given that the quiescent wind values of GJ 436 and Proxima Cen derived in \citet{2017MNRAS.470.4026V} and \citet{2001ApJ...547L..49W} are below our minimum $\dot{M}$ limit,  it is unlikely that radio emission from GJ 436 b and Proxima b would be detectable with present-day instruments. Detectable emission from these planets could be generated if, for example, these planets interact with coronal mass ejections, which momentarily increase stellar wind densities and velocities. Figure \ref{fig.min_wind} shows other exoplanets, such as GJ 674 b, GJ 876 b and YZ Cet b, that present good prospects for radio detection, provided that their quiescent wind have $\dot{M} u_{\rm sw}^{3} \gtrsim 1.8\e{-4}$.

Given that the radio flux is sensitive to the local density and velocity of the stellar wind, we might not be able to decouple these two quantities without additional hypotheses (or additional observations), but certainly this would bring us one step closer to observationally constraining the winds of M dwarf stars. And once more exoplanets could be useful `wind-ometers', i.e., tools to probe winds of their host stars \citep{2017MNRAS.470.4026V}.
 
\section*{Acknowledgements}
AAV acknowledges funding from the Irish Research Council Consolidator Laureate Award 2018 and thanks Dr Blakesley Burkhart for promptly clarifying details of their models. This research has made use of the NASA Exoplanet Archive, which is operated by the California Institute of Technology, under contract with the National Aeronautics and Space Administration under the Exoplanet Exploration Program. We thank the anonymous reviewer for their constructive comments.

\let\mnrasl=\mnras
\input{dM_radio.bbl}
\bsp
\label{lastpage}

\end{document}